\documentstyle[aps,prc,epsf,twocolumn]{revtex}
\begin{document}
\title{A Bethe-Salpeter model for light mesons: spectra and decays}
\author{C.R.M\"unz, J.Resag, B.C.Metsch, H.R.Petry}
\address{Institut f\"ur Theoretische Kernphysik,\\
         Universit\"at Bonn, Nussallee 14-16, 53115 Bonn, FRG}
\date{\today}
\maketitle

\begin{abstract}
The spectra and electroweak decay properties of light mesons are
analyzed within the framework of the instantaneous Bethe-Salpeter
equation. The interaction kernel comprises alternative spin-structures
for a parameterization of confinement and a residual quark-antiquark
interaction based on instanton effects. It is shown that only with a
vector confinement the parameters can be chosen such as to yield an
excellent description of the light pseudoscalar and vector mesons
including weak and two photon decays. However it is found that it is
not possible to reconcile this with the Regge behavior of higher
lying meson states with the same parameter set.
\end{abstract} \pacs{}

\narrowtext

\section{Introduction} \label{I}
In a previous paper \cite{RMMP} it has been shown that the
Bethe-Salpeter(BS) equation
with an instantaneous interaction (Salpeter equation) provides a suitable
framework for relativistic quark models of light mesons.
The use of the Salpeter equation has several advantages compared
to other treatments, i.e.
\begin{itemize}
\item
The relativistic kinematics of the quarks is treated correctly;
\item
The amplitudes have the correct relativistic normalization;
\item
The lower component \(\Phi^{--}\) of the Salpeter amplitude
is determined dynamically.
\end{itemize}
On the other hand the practical advantages of a nonrelativistic treatment
are also present, i.e.
\begin{itemize}
\item
The Salpeter equation can be formulated as an eigenvalue problem
\({\cal H}\psi=M\psi\) for the mass \(M\) of the bound state.
\item
One can define a (not positive definite)
scalar product \( \langle \psi_a | \psi_b \rangle \)
for the Salpeter amplitudes.
\item
The Salpeter operator \({\cal H}\) is selfadjoint with respect
to this scalar product.
\end{itemize}
In \cite{RMMP} we have developed a numerical scheme
which enables the calculation of meson mass spectra and the
corresponding BS-amplitudes. Furthermore we have given formulas
to compute some important electroweak meson decay widths.

In the present paper we shall apply this method to an explicit quark
model for light mesons. Our main concern in this context will be
whether a realistic description of deeply bound states like the pion
is compatible with a reasonable description of confinement.

Let us first give a list of the main features of light mesons
that will be considered in the following:
\begin{itemize}
\item
the low masses of \(\pi\) and \(K\)
\item
the weak decay constants \(f_{\pi}\) and \(f_K\)
\item
the decays \(\pi^0,\,\eta,\,\eta' \rightarrow 2\,\gamma\)
\item
the masses and the flavor mixing coefficients of \(\eta\) and \(\eta'\)
\item
the masses and the leptonic decay widths for the
\(\rho,\,\omega\) and \(\phi\) mesons
\item
the Regge behavior \(M^2 \sim J\)
\end{itemize}
As far as we know there presently
is no model that can describe all these features
in a consistent way. On the one hand there is the nonrelativistic quark
model that gives a reasonable description of the mass spectra \cite{Bla},
but that completely fails in describing the decay widths of the deeply
bound states like the pion (see Sec.\ref{V1}). On the other hand
there are models like the Nambu Jona-Lasinio model \cite{NJL,Kli,Wei}
that are based on the chiral symmetry of QCD for vanishing current
quark masses. This model leads to a good description for the
\(\pi, K, \eta\) and \(\eta'\) mesons, but
higher angular momenta states or radial excitations
cannot be described since confinement is ignored.

An attempt to arrive at a more complete description
based on the Salpeter equation
has recently been given by J.F.Laga\"e \cite{La}. His results
show the difficulty of finding a suitable ansatz for the confining
interaction kernel: he rejects the hypothesis of an instantaneous
scalar confining kernel.

The model we present in the following is based on a linear scalar
or alternatively vector
confining kernel combined with an effective interaction computed
by 't Hooft from instanton effects in QCD \cite{Hoo,SVZ,Pet}.
A nonrelativistic version of this interaction has already
lead to good results for the meson and baryon mass spectra
\cite{Bla}. We therefore feel encouraged to test this ansatz
in the relativistic Bethe-Salpeter framework.

The paper is organized as follows:
In Sec.\ref{II} we briefly summarize the main features of
the Salpeter equation as given in ref.\cite{RMMP}.
The explicit form of the confining BS-kernel and of the
't Hooft kernel is given in Sec.\ref{III}.
In Sec.\ref{IV} we present the calculated meson mass spectra
and obtain the pion and kaon decay constants \(f_{\pi},\,f_K\),
the decay width into two photons for the \(\pi^0,\,\eta\)
and \(\eta'\) mesons and the leptonic widths for the vector
mesons \(\rho,\,\omega\) and \(\phi\). We will compare
these decay widths to corresponding nonrelativistic results
in Sec.\ref{V1} using
the wave function of ref.\cite{Bla}, which shows the impressive
improvement due to the relativistic treatment of the quarks
compared to the nonrelativistic potential model.
Finally we give some concluding remarks in Sec.\ref{VI}.

\section{The Salpeter equation} \label{II}
For an instantaneous BS-kernel and free quark propagators
the \(p^0\) integrals in the BS-equation can be computed
analytically in the rest frame of the bound state with mass \(M\).
The result is known as the Salpeter equation which reads
\begin{eqnarray}
\Phi(\vec{p}) &=&
\int\!\! \frac{d^3p'}{(2\pi)^3} \,
\frac{ \Lambda^-_1(\vec{p}) \, \gamma^0 \,
[ V(\vec{p},\vec{p}\,') \, \Phi(\vec{p}\,') ]
\, \gamma^0 \, \Lambda^+_2(-\vec{p})}
{M+\omega_1+\omega_2}
 \nonumber \\
 &-&
\int\!\! \frac{d^3p'}{(2\pi)^3}\,
\frac{\Lambda^+_1(\vec{p})\,\gamma^0\,
[(V(\vec{p},\vec{p}\,')\,\Phi(\vec{p}\,')]
\,\gamma^0\,\Lambda^-_2(-\vec{p})}
{M-\omega_1-\omega_2}
\label{BSe}
\end{eqnarray}
with \(\omega_i=\sqrt{\vec{p}\,^2+m_i^2}\), \(\;\;
\Lambda^{\pm}_i(\vec{p}) = (\omega_i \pm H_i(\vec{p}))/(2\omega_i)
\) and
\(H_i(\vec{p})=\gamma^0(\vec{\gamma}\vec{p}+m_i)\).

This equation can be reformulated
into an eigenvalue problem for the bound state
mass \(M\). With the definitions
\begin{eqnarray}
\psi(\vec{p}) &:=& \Phi(\vec{p})\,\gamma^0 \label{17}\\
 {} [W(\vec{p},\vec{p}\,')\, \psi(\vec{p}\,')] &:=&
\gamma^0\,[V(\vec{p},\vec{p}\,')\, \Phi(\vec{p}\,')]
\end{eqnarray}
the Salpeter equation can now be written as
\begin{eqnarray}
{\cal H}\psi &=& M\,\psi \;\;\;\;\mbox{with} \nonumber \\
{\cal H}\psi
 &=& H_1 \psi - \psi H_2 \nonumber \\ &-&
  \int' \left\{ \Lambda_1^+ \,[W \psi')]\,\Lambda_2^-
 -             \Lambda_1^- \,[W \psi')]\,\Lambda_2^+ \right\}
\end{eqnarray}
with all quantities depending on \(\vec{p}\) or \(\vec{p}\,'\)
(indicated with a prime)
and the notation \(\int'=\int\,d^3p'/(2\pi)^3\).

One can now define a (not positive definite)
scalar product for amplitudes \(\psi_1=\Phi_1\gamma^0\)
and \(\psi_2=\Phi_2\gamma^0\) as
\begin{equation}
\left\langle \psi_1 \right.\left| \psi_2 \right\rangle
= \int\,\mbox{tr}\,\left(
    \psi_1^{\dagger}\,\Lambda_1^+\,
    \psi_2\,\Lambda_2^- -
    \psi_1^{\dagger}\,\Lambda_1^-\,
    \psi_2\,\Lambda_2^+ \right)
\label{scprd}
\end{equation}
The normalization condition for solutions of the
Salpeter equation is then given as
\( \langle \psi | \psi \rangle = (2\pi)^2\,2M \).
If one considers amplitudes satisfying
\(\Lambda_1^+ \psi \Lambda_2^+ = \Lambda_1^- \psi \Lambda_2^- = 0\)
one easily finds
\begin{eqnarray}
\left\langle \psi_1 \right.\left|
\,{\cal H}\, \psi_2 \right\rangle
&=& \int\,(\omega_1+\omega_2)\,\mbox{tr}\,
\left(\psi_1^{\dagger}\,\psi_2\right) \nonumber \\ &-&
\int\,\int'\, \mbox{tr}\,
\left(\psi^{\dagger}_1\,W\psi_2'\right) \label{expH}
\end{eqnarray}
For most kernels of physical interests \({\cal H}\) is selfadjoint
with respect to the scalar product given above which implies that
bound state masses \(M\) are real for eigen functions \(\psi\)
with nonzero norm
and that amplitudes \(\psi_1\) and \(\psi_2\) corresponding to different
eigenvalues \(M_1 \not= M_2^*\) are orthogonal, i.e.
\(\left\langle \psi_1 \right.\left|\,\psi_2 \right\rangle = 0\).

\section{The Bethe-Salpeter kernel} \label{III}
\subsection{Confinement} \label{IIIA}
Up to now the confining interaction of QCD is only known in the static
limit of heavy quarks. In this limit it has been shown \cite{LSG,Gro}
that the static potential between quarks is of the form
\(V_C(r)= a_c+b_cr+W \) where \(W\) denotes the relativistic corrections
of the order \(p^2/m^2\). As stated by Gromes \cite{LSG} it is still
an open question whether an additional \(1/r\) term should also be included
into the static confining potential. Usually one concludes from the sign
of the spin-orbit coupling term in \(W\) that the confining
\(q\bar{q}\)-interaction behaves like a Lorentz scalar.

The problem for the case of light quarks is that up to now there is
no unambiguous extension of the confining potential beyond the
static limit. Especially there is no prescription on how to extend
it to a noninstantaneous form. Naive noninstantaneous extensions
fail as has been shown by S.N.Biswas et al. \cite{Bis} for the
harmonic oscillator BS-kernel \(V(x)=-bx^2=b\,(\vec{x}\,^2-(x^0)^2)\)
that yields only a continuous spectrum. Similar results
are to be expected for other kernels like \(1/q^4\). Because
of these difficulties the only way we see at the moment is to
parameterize confinement as an instantaneous interaction kernel.

The sign of the LS-term in the static limit would be compatible with
a scalar confinement kernel. However some authors \cite{La,Ga} have shown
that the linear Regge behavior \(M^2 \sim J\) is lost
for this choice, since bound state masses come out too
small for higher angular momenta. This is due to the relativistic
corrections to the static potential and becomes more problematic
with decreasing quark masses.
For a vector confining kernel this problem almost disappears,
but in the static limit
a vector kernel leads to an LS-term which has the wrong sign.
To our knowledge there presently is no convincing parameterization
for the confining kernel that exhibits both features, i.e. leads to linear
Regge trajectories and yields a spin-orbit term with the correct sign.

In the following we will analyze both spin structures
for the confinement, i.e. a scalar \(1 \otimes 1\) and a
vector \(\gamma^0 \otimes \gamma_0\) interaction.
In the rest frame of the bound state the corresponding BS-kernels in
eq.(\ref{BSe})
are parameterized as
\begin{eqnarray}
\left[V_C^V(\vec{p},\vec{p}\,')\,\Phi(\vec{p}\,')\right]
&=& -{\cal V}_C((\vec{p}-\vec{p}\,')^2)\,\gamma^0\,\Phi(\vec{p}\,')\,\gamma^0
\label{vector} \\
\left[V_C^S(\vec{p},\vec{p}\,')\,\Phi(\vec{p}\,')\right]
&=& \;\;\;{\cal V}_C((\vec{p}-\vec{p}\,')^2)\,\Phi(\vec{p}\,')
\label{scalar}
\end{eqnarray}
where \({\cal V}_C\) is a scalar function which has the fourier
transform \({\cal V}_C^F(r) = a_c+b_c r\).

\subsection{'t Hooft interaction} \label{IIIB}
\subsubsection{The 't Hooft lagrangian} \label{IIIB1}
It is already known from nonrelativistic potential models that
the masses of the scalar and pseudoscalar mesons \(\pi,\,K,\,\eta,\,\eta'\)
cannot be described with a confining potential alone. The usual extension
would be to add another contribution to the interaction kernel
that comes from a One Gluon Exchange (OGE). This works quite well
for heavy quarkonia \cite{Ga,Bey}, but it is highly questionable
for light mesons where perturbation theory cannot be expected to work.
Since OGE leads to a flavor independent interaction kernel one thus
obtains degenerate \(\pi\) and \(\eta\) mesons in clear contradiction
to the experimental mass values
\(m_{\pi}=140\,MeV\) and \(m_{\eta}=549\,MeV\).
In order to cure this discrepancy one would have to take into account
higher order diagrams.

However there is another QCD based candidate for a residual
\(q\bar{q}\)-interaction computed by 't Hooft and others from
instanton effects \cite{Hoo,SVZ,Pet} which leads to good results
for meson and baryon mass spectra within a nonrelativistic
potential model \cite{Bla}.

Instantons are special solutions of the classical nonabelian
Yang-Mills equations in Euclidian space. They are peaked both
in space and imaginary time having a finite extension \(\rho\).
Since they cannot be deformed continuously into classical solutions
corresponding to gluon fields they lead to an effective interaction between
quarks that is not covered by perturbative gluon diagrams.
This interaction leads to spontaneous breaking of chiral
symmetry as can be seen by normal ordering of the underlying
Lagrangian (see appendix \ref{thoapp}).
The normal ordered Lagrangian takes the form
\begin{equation}
{\cal L} = k + \sum_{j=1}^3\,:\!(i\bar{q}_j\gamma^{\mu}\partial_{\mu}q_j
- m_j\,\bar{q}q)\!: + \Delta{\cal L}{(2)}  + \Delta{\cal L}{(3)}
\label{normord}
\end{equation}
where \(k\) is an inessential constant that renormalizes the vacuum energy.
\(\Delta{\cal L}{(2)}\) and \(\Delta{\cal L}{(3)}\) are two and three body
terms and \(m_j=m_j^0+\Delta m_j\) is the effective constituent quark mass.
In the following we will consider these terms.

\subsubsection{The constituent quark mass}
The contribution \(\Delta m_j\) to the effective constituent
quark mass \(m_j=m_j^0+\Delta m_j\) is given by
\begin{eqnarray}
  \Delta m_{n} &=& \int \limits_0^{\rho_{c}}
                   d\rho \frac{d_0(\rho)}{\rho^5}
                         \frac{4}{3} \pi^2 \rho^3
                         \left( m_{n}^0\rho - \frac{2}{3}\pi^2\rho^3
                                \langle \bar{q}_nq_n \rangle \right)
\nonumber \\ && \;\;\;\;\;\;\;\cdot
                         \left( m_{s}^0\rho - \frac{2}{3}\pi^2\rho^3
                                \langle \bar{q}_sq_s \rangle \right)
\nonumber \\
  \Delta m_{s} &=& \int \limits_0^{\rho_{c}}
                   d\rho \frac{d_0(\rho)}{\rho^5}
                         \frac{4}{3} \pi^2 \rho^3
                         \left( m_{n}^0\rho - \frac{2}{3}\pi^2\rho^3
                                \langle \bar{q}_nq_n \rangle \right)^2
\label{gapeq}
\end{eqnarray}
where the instanton density for three colors and three flavors reads
\cite{Shu}
\begin{equation}
  d_0(\rho) = (3.63 \cdot 10^{-3}) \left( \frac{8\pi^2}{g^2(\rho)} \right)^{6}
              \exp{ \left( - \frac{8\pi^2}{g^2(\rho)} \right) }
\end{equation}
with
\begin{equation}
  \left( \frac{8\pi^2}{g^2(\rho)} \right) =
                9 \ln{ \left( \frac{1}{\Lambda_{_{QCD}}\,\rho} \right) } +
  \frac{32}{9}\ln{\ln{ \left( \frac{1}{\Lambda_{_{QCD}}\,\rho} \right) }}
\end{equation}
within two loop accuracy \cite{SG}.
Here \(\Lambda_{_{QCD}}\) is the QCD scale parameter.
The integration over the instanton size \(\rho\) has to be
carried out up to a cutoff value \(\rho_c\) where the
\(ln\,ln\) term coming from the two loop correction is still
small compared to the \(ln\) term.

Equations (\ref{gapeq}) are usually called Gap equations.
They describe the generation of a dynamical quark mass due to the interaction
with the negative Dirac see. In our model the constituent quark masses \(m_j\)
will be used as free parameters that are fitted to the experimental data.
It will be checked in the end (see Sec.\ref{V2})
whether the obtained quark masses are compatible
with common values for the quark condensates present in \(\Delta m_j\)
assuming that the confinement interaction does not contribute essentially
to the process of chiral symmetry breaking.

\subsubsection{The two body interaction}
The two body term reads (see appendix \ref{thoapp})
\begin{eqnarray}
\lefteqn{ \Delta{\cal L}{(2)} = -\frac{3}{16}
 \sum_{i} \sum_{kl} \sum_{mn}\, g_{\mbox{eff}}(i)\,
      \varepsilon_{ikl}\varepsilon_{imn} }
      \nonumber \\ & &
      \Bigl\{
        \,: q_k^{\dagger} q_l^{\dagger}\,
      \bigl[ \gamma_0 \! \cdot \! \gamma_0
           + \gamma_0 \gamma_5 \! \cdot \! \gamma_0 \gamma_5
      \bigr]
      \,(2{\cal P}^C_{\bar{3}}+{\cal P}^C_6)\,
        q_m q_n :\,
      \Bigr\} \label{dl2}
\end{eqnarray}
where the effective coupling constants are given as
\begin{equation}
    g_{\mbox{eff}}(i) = \int \limits_0^{\rho_{c}}
      d\rho \frac{d_0(\rho)}{\rho^5}
      \left( \frac{4}{3} \pi^2 \rho^3 \right)^2
      \left( m_{i}^0\rho
           - \frac{2}{3}\pi^2\rho^3 \langle \bar{q}_iq_i \rangle \right)
\label{geff} \end{equation}
and the tensor notation
\begin{equation}
q^{\dagger} q^{\dagger}\,(A \cdot B)\, q\, q :=
\sum_{i,j}\,\sum_{k,l}\,
q_{i}^{\dagger} q_{j}^{\dagger}\,A_{ik} \cdot B_{jl}\, q_{k} q_{l}
\end{equation}
has been used for Dirac and color indices.
This representation explicitly shows the antisymmetric flavor
dependence of the interaction.
The color sextett and antitriplett
projection matrices are given by
\begin{eqnarray}
{\cal P}^C_6 &=& \frac{1}{2} \,(1^C+\Pi^C)
 = \frac{2}{3}\,1^C +
     \frac{1}{4}\,\vec{\lambda} \!\cdot\! \vec{\lambda} \\
{\cal P}^C_{\bar{3}} &=& \frac{1}{2} \,(1^C-\Pi^C)
 = \frac{1}{3}\,1^C -
     \frac{1}{4}\,\vec{\lambda} \!\cdot\! \vec{\lambda}
\end{eqnarray}
so that
\begin{equation}
2{\cal P}^C_{\bar{3}}+{\cal P}^C_6 = \frac{1}{2}\,(3\,1^C - \Pi^C)
= \frac{4}{3}\,1^C - \frac{1}{4}\,\vec{\lambda} \!\cdot\! \vec{\lambda}
\end{equation}
where $\Pi^C$ is a color exchange matrix defined as
\(
\Pi^C_{ij,kl} = \delta_{il}\,\delta_{jk}
\)
and \(\lambda^a \;(a=1,\ldots,8)\) are the color matrices.
Just like the quark masses also the coupling constants will be treated
as free parameters in our model.

\subsubsection{The three body interaction}
After a lengthy calculation the three body force can finally be written
in the form
\begin{eqnarray}
  \Delta{\cal L}{(3)} & = & \frac{27}{80} g_{\mbox{eff}}^{(3)}
      \Bigl\{
        \,:\, q^{\dagger} q^{\dagger} q^{\dagger}
      \nonumber \\ & &
      \bigl[ \gamma_0 \! \cdot \! \gamma_0 \! \cdot \! \gamma_0
           + \gamma_0 \gamma_5 \! \cdot \! \gamma_0
             \gamma_5 \! \cdot \! \gamma_0 \nonumber \\ &&\;
           + \gamma_0 \gamma_5 \! \cdot \! \gamma_0
             \! \cdot \! \gamma_0 \gamma_5
           + \gamma_0 \! \cdot \! \gamma_0 \gamma_5
             \! \cdot \! \gamma_0 \gamma_5
      \bigr]
      \nonumber \\ & &
      {\cal P}^F_1 (2{\cal P}^C_{10}+5{\cal P}^C_8)\,
        q q q \,:\,
      \Bigr\}
\end{eqnarray}
where ${\cal P}^F_1$ is the projector onto a three-particle flavor
singulett state, ${\cal P}^C_{10}$ and ${\cal P}^C_8$ are projectors
onto the color decuplett and the color octett. The effective three-body
coupling constant is given by
\begin{equation}
g_{\mbox{eff}}^{(3)}=\int_0^{\rho_c}\,
d\rho\,\frac{d_0(\rho)}{\rho^5}\,\left(\frac{4}{3}\pi^2\rho^3\right)^3
\end{equation}
Obviously this three body force does not contribute to \(q\bar{q}\)-states
and to colorfree \(qqq\)-states.

\subsubsection{The 't Hooft kernel}
\begin{figure}[ht]
  \centering
  \leavevmode
  \epsfxsize=0.25\textwidth
  \epsffile{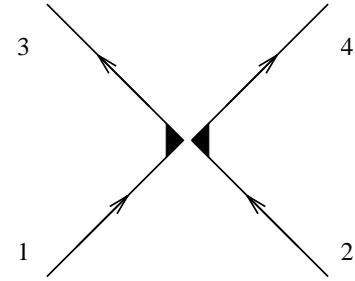}
\vspace{0.5cm}
\caption{Instanton induced interaction vertex corresponding to \(\Delta {\cal
L}_2\).}
\label{figthovertex}
\end{figure}

\begin{figure}[ht]
  \centering
  \leavevmode
  \epsfxsize=0.40\textwidth
  \epsffile{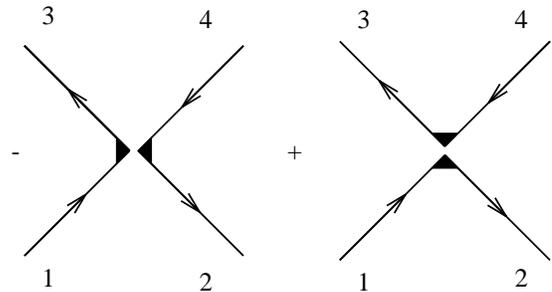}
\vspace{0.5cm}
\caption{Instanton induced interaction vertices for \(q\bar{q} \rightarrow
q\bar{q}\).}
\label{figthoqqbar}
\end{figure}

In order to distinguish different indices in the following
we will use the notation \(s_i\) for Dirac indices,
\(c_i\) for color indices and \(f_i\) for flavor indices.
The vertex corresponding to \(\Delta {\cal L}_2\) is
shown in Fig.\ref{figthovertex}.
The vertex reads
\begin{eqnarray}
&&(-i)\,G_{f_3f_4,f_1f_2}\, (1_{s_3s_1}1_{s_4s_2} +
\gamma^5_{s_3s_1}\gamma^5_{s_4s_2})
\nonumber \\&&
(4/3\;1_{c_3c_1}\,1_{c_4c_2}
 - 1/4\;\vec{\lambda}_{c_3c_1}\vec{\lambda}_{c_4c_2})
\end{eqnarray}
with the definition
\begin{equation}
G_{f_3f_4,f_1f_2} := \frac{3}{8}\,
\sum_{f_5} \,g_{\mbox{eff}}(f_5)\;\epsilon_{f_5f_3f_4}\,\epsilon_{f_5f_1f_2}
\label{gff}
\end{equation}
For the \(q\bar{q} \rightarrow q\bar{q}\) amplitude
we have to consider the two diagrams given in Fig.\ref{figthoqqbar}.
In a meson the quark and the antiquark are in a color singulett state
represented by the matrix
\( \chi^C_{c_1c_2} = \delta_{c_1c_2}\,/\sqrt{3} \).
The color matrix elements for the two vertices are then given by
\begin{eqnarray}
\sum_{c_1c_2}\sum_{c_3c_4}\,(\chi^C_{c_3c_4})^*\,\frac{1}{2}
\,(3\,1_{c_3c_1} 1_{c_2c_4} - \Pi_{c_3c_2,c_1c_4})\,\chi^C_{c_1c_2} &=& 0\\
\sum_{c_1c_2}\sum_{c_3c_4}\,(\chi^C_{c_3c_4})^*\,\frac{1}{2}
\,(3\,1_{c_3c_4} 1_{c_2c_1} - \Pi_{c_3c_2,c_4c_1})\,\chi^C_{c_1c_2} &=& 4
\end{eqnarray}
so that only the second vertex
in Fig.\ref{figthoqqbar} contributes to the interaction kernel.
The effective 't Hooft interaction vertex between
\(q\bar{q}\) color singulett states is then given by
\begin{equation} (-4i)\,G_{f_2f_3,f_1f_4} (1_{s_3s_4}1_{s_2s_1}
 + \gamma^5_{s_3s_4}\gamma^5_{s_2s_1})
\end{equation}
Since we assume \(SU(2)\)-flavor invariance of the interaction
we set
\begin{equation} g := \frac{3}{8}g_{\mbox{eff}}(s) \;\;\;,\;\;\;
   g':= \frac{3}{8}g_{\mbox{eff}}(n) \label{ggp}\end{equation}
where \(s\) stands for strange and \(n\) stands
for nonstrange (u,d) flavor.
The results for \(G_{f_2f_3,f_1f_4}\) are given in Tab.\ref{tab1}.
Tab.\ref{tab2} shows the matrix elements of \(G_{f_2f_3,f_1f_4}\) for the
flavor functions
\begin{eqnarray}
\pi^0 &=&  (u\bar{u}-d\bar{d})/\sqrt{2} \\
\eta_n &=& (u\bar{u}+d\bar{d})/\sqrt{2} \\
\eta_s &=& s\bar{s}
\end{eqnarray}
{}From the vertex we can extract the lowest order contribution
of the 't Hooft interaction to the BS-kernel as
\begin{eqnarray}
\lefteqn{[V_T(\vec{p},\vec{p}\,')\,\Phi(\vec{p}\,')]_{f_1f_2}
 = 4\,\sum_{f_1'f_2'}
\,G_{f_2f_1',f_1f_2'} }  \nonumber \\
& &
\left[ 1\,\mbox{tr}\,\left(\Phi_{f_1'f_2'}(\vec{p}\,')\right) +
  \gamma^5\,\mbox{tr}\,\left(\Phi_{f_1'f_2'}(\vec{p}\,')\,
  \gamma^5\right)\,\right]
\label{thokern} \end{eqnarray}
As shown in the appendix this interaction only acts on the scalar and
pseudoscalar
mesons \(J^{\pi_P}=0^{\pm}\). For the pseudoscalar mesons it is
attractive for the \(\pi\) with
a coupling constant \(g\) and for the \(K\) with a coupling constant \(g'\).
For the \(\eta\) and \(\eta'\)
the interaction leads to mixing of nonstrange and strange flavor
amplitudes. The effective sign of the interaction is reversed
for the scalar mesons thus being repulsive for the \(a_0\).
Note that in the nonrelativistic limit the 't Hooft interaction
only acts on the pseudoscalar states \(J^{\pi_P}=0^-\).
The 't Hooft kernel as it stands represents a pointlike
interaction that has to be regularized. Following ref.\cite{Bla}
we do this by multiplying the kernel with a regularizing Gaussian function
\begin{equation}
{\cal V}_{\mbox{reg}}(q)=e^{-\frac{1}{4}\Lambda^2q^2}
\end{equation}
with \(\vec{q}=\vec{p}-\vec{p}\,'\) and \(q=|\vec{q}|\).
In coordinate space this choice corresponds to replacing
the \(\delta(\vec{r})\) function by
\begin{equation}
{\cal V}_{\mbox{reg}}^F(r)=\frac{1}{(\Lambda \sqrt{\pi})^3}\,
             e^{-\frac{r^2}{\Lambda^2}} \label{regtho}
\end{equation}
which introduces a finite effective range \(\Lambda\).

\section{Results and discussion}
\label{IV}
\subsection{Models and Parameters}
\label{MoPa}
The main concern of our work was to see whether we can obtain
a consistent description of
a) the masses and decays of the low lying pseudoscalar and vector mesons
and b) confinement reflected e.g. by the Regge
trajectories.

For this purpose we investigate two different models of the
confinement kernel: 1)  a vector \(\gamma^0\otimes\gamma^0\)- and 2) a
scalar \( 1\otimes 1\)-structure.

The parameters used are the nonstrange and strange quark masses
\(m_n\) and \(m_s\), the offset \(a_c\) and slope \(b_c\) of the
confinement interaction, the two coupling constants g,
g' and the effective range \(\Lambda\) of the residual instanton
induced interaction. So the total number of parameters amounts to
seven.

We used two sets of parameters in the vector confinement case:
Model V1 was tuned to reproduce the masses and decays of the low lying mesons.
We therefore used a small nonstrange quark mass \(m_n\), as
the correct description of the pseudoscalar decays depends
essentially on this quantity. Given this mass we had to take a
moderate confinement slope to reproduce the decays of the vectormesons.
The offset \(a_c\) was fixed by the \(\rho\)-mass and \(m_s\)
by the \(K^*\)-mass.
Finally g and g' were fixed by the masses of the
pseudoscalars \(\pi , \eta\) and K. In Model V2 we used a
larger nonstrange mass of \(m_n\) of about 1/3 of the nucleon mass,
which is a value common to Nonrelativistic Quark Models. The
aim of this parameter set was to obtain a good description of the
Regge trajectories and the higher lying resonances.

Finally in Model S we investigated a scalar confinement
provided with the same quark mass \(m_n\) as in V2.
For this kernel it turned out
that our method of solving this type of BS equation works reasonably
for higher angular momenta only if
the fraction of the confinement slope \(b_c\) and the quark mass
\(m_q\) is sufficiently small.
The parameters for the three model are listed in Tab.\ref{param}.

\subsection{Mass spectra}
\label{Msspc}
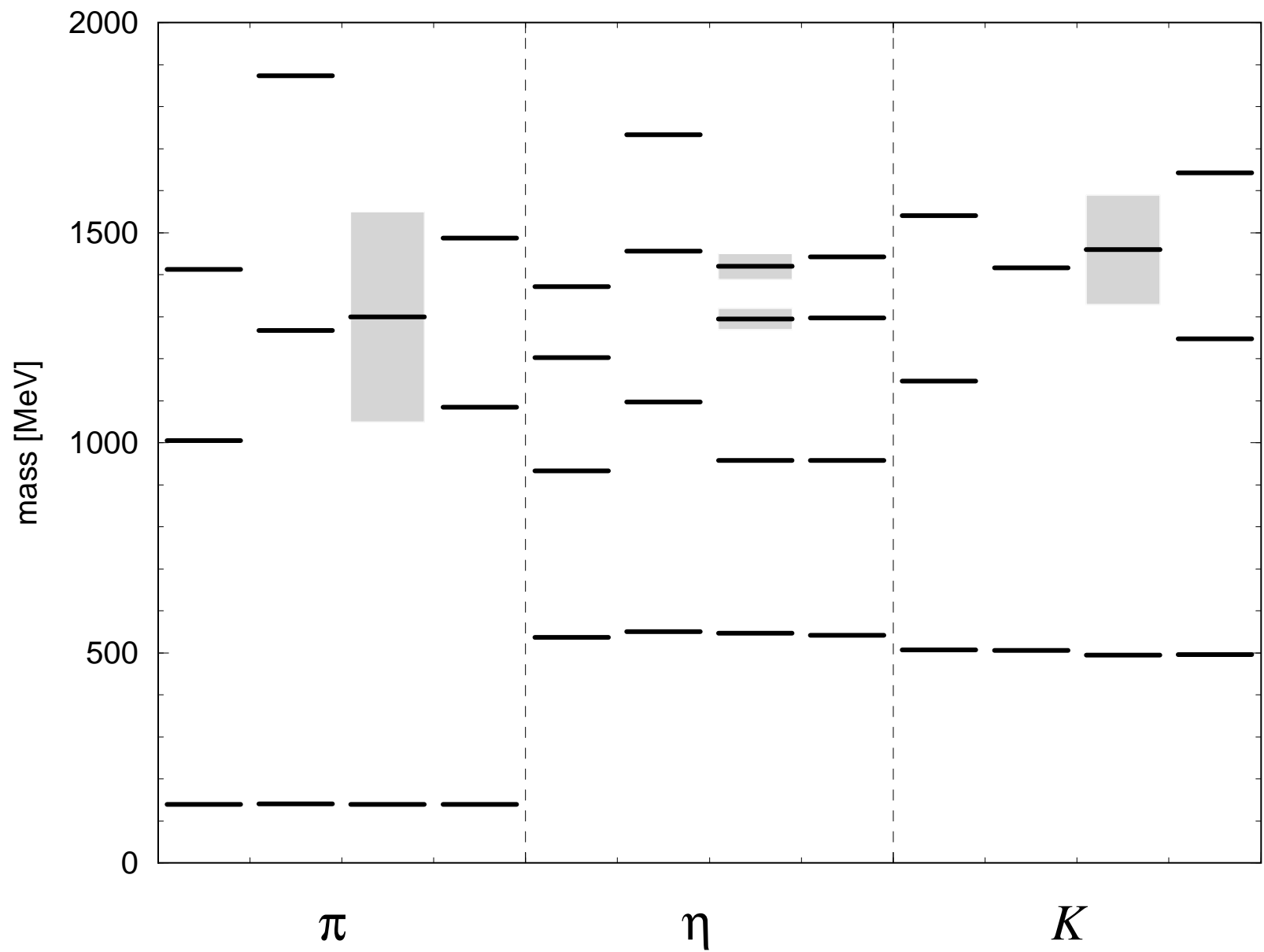
\begin{figure}
\caption{Mass spectra of the pseudoscalar mesons.
The columns for each meson correspond (from the left) to model V1,
model V2, experiment~\protect\cite{PDG} and model S. The shaded areas (3rd
column)
indicate the experimental full width of the meson.}
\label{spectra1}
\end{figure}

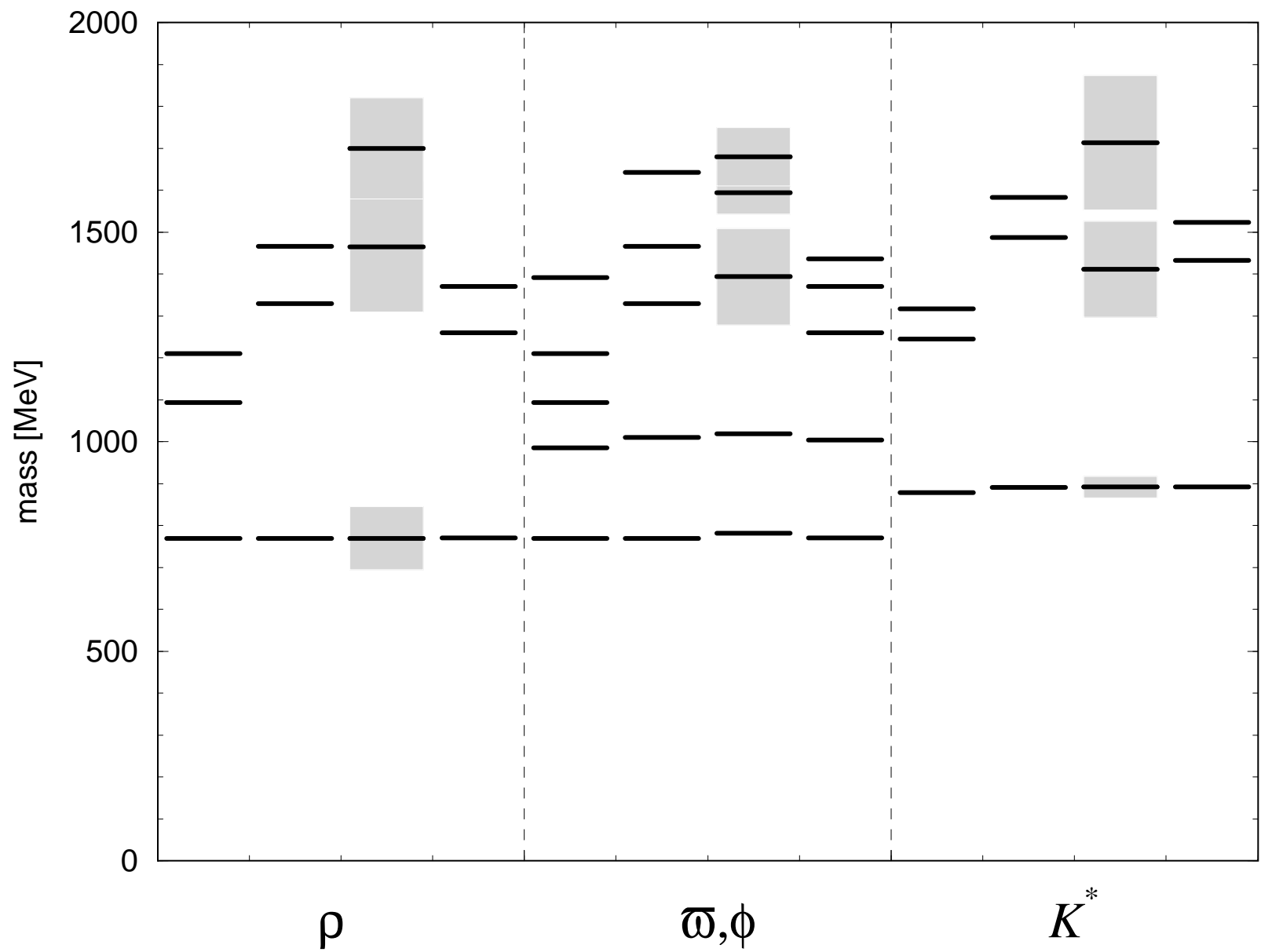
\begin{figure}
\caption{Mass spectra of the vector mesons (see also caption to
         Fig.~\protect\ref{spectra1}).}
\label{spectra2}
\end{figure}

The quality of the mass spectra is different for the three models, as
different priorities led us to the parameters.
Common to all three models is an overall agreement of the masses of
the pseudoscalars \( \pi ,K, \eta, \eta' \) and the vector mesons \( \rho,
\omega, \phi, K^* \). The spectra for these meson are compared to
experiment in Figs. \ref{spectra1},\ref{spectra2}.
The 't Hooft interaction leads to the correct
splitting of \(\pi,\eta\) and \(\eta'\) mass.
In contrast with experiment, where the \(a^0\) and \(f^0\) are nearly
degenerate, we
obtain a large splitting due to the instanton induced interaction of
several hundred MeV (compare Tab.\ref{spor}).
For a vector type kernel with positive spin orbit
splitting this leads an enormous attraction: although there is a scalar state
at roughly 1GeV, we also find a state with imaginary
mass and zero norm.  The physical interpretation of this phenomenon
however is not
clear. For a scalar confinement this effect is compensated by the
negative spin orbit splitting. In the following we will discuss the
differences of the three parameter sets.

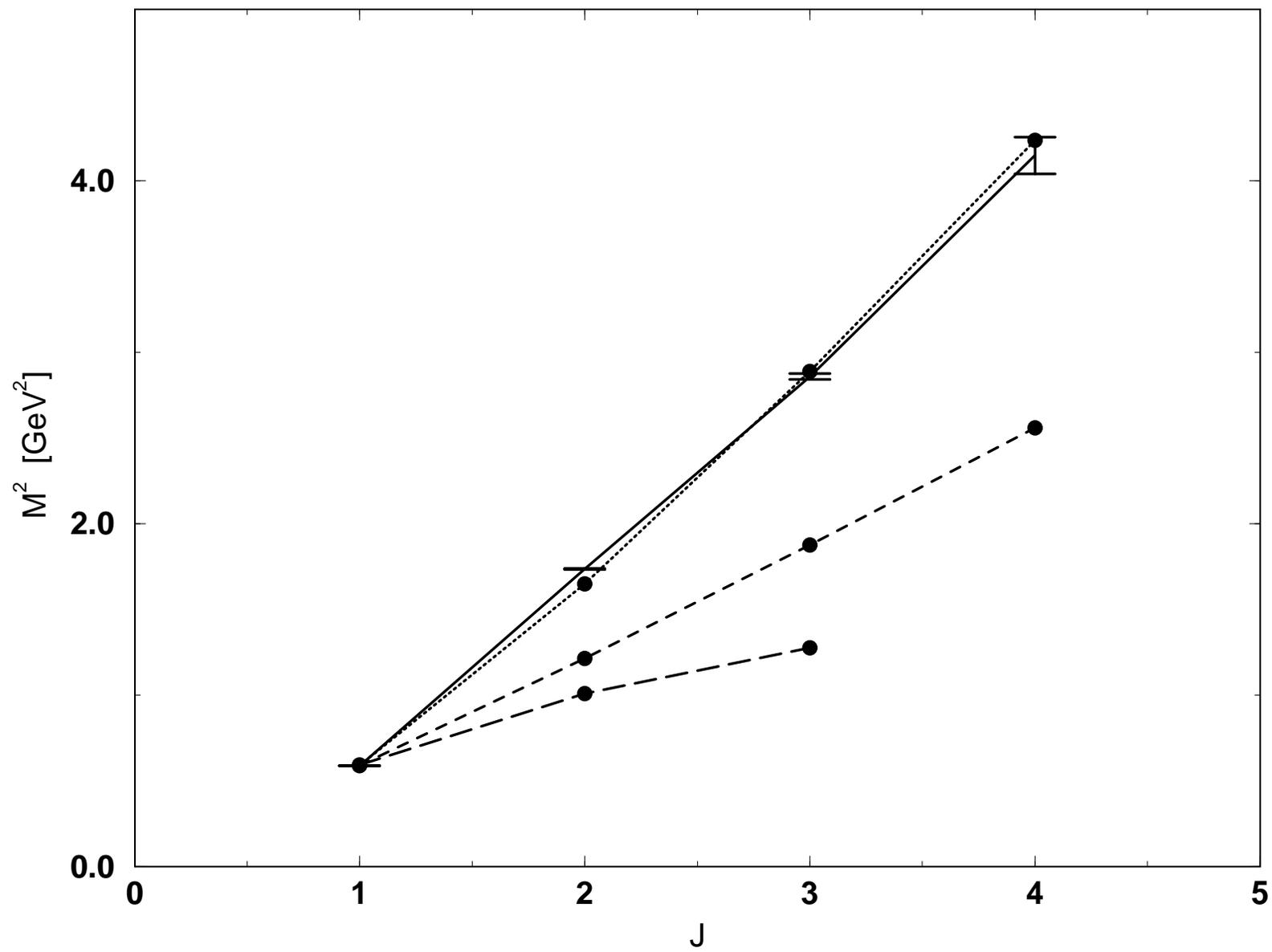
\begin{figure}
\caption{Regge trajectory for the isovector mesons with \(S=1\).
The solid line shows the experimental
masses for \(\rho,\,a_2,\,\rho_3,\,a_4\) ~\protect\cite{PDG} where the errorbar
gives the experimental error for the resonance position.
The short dashed line corresponds to the calculated masses
of model V1, the dotted line to model V2 and the long dashed line
to the scalar confinement (model S).}
\label{Regge}
\end{figure}

Since in model V1 we used a small confinement slope to reproduce
the decays of the vectormesons, the calculated Regge trajectory is too
flat (see Fig.\ref{Regge}).
The spin orbit splitting between the \(1^{++}\) and \(2^{++}\) mesons
in our model is purely due to the confining interaction. In both models
with vector kernel it is of order of 200 MeV exceeding the
experimental mass difference, which is in fact rather small (see
Tab.\ref{spor}).

In model V2 with large
quark masses and large confinement slope \(b_c\) we obtain a good
description of the masses of all mesons comparable to the results in
nonrelativistic calculations. However \(b_c\) has to be much larger in
the BS framework, as obviously the kinetic energy is overestimated in
nonrelativistic calculations. The Regge trajectories representing the
confinement property are well reproduced (Fig.\ref{Regge}). For the
spin orbit splitting the same remarks as for V1 apply.

Finally model S shows a reasonable description of the
ground states of the pseudoscalar and vector mesons.
For the states with large angular momentum the method of solving the
BS equation by a basis expansion does not lead to convergent solutions
with positive norm. With increasing dimension of the basis the smallest
positive
eigenvalue decreases until it becomes imaginary (a similar
problem arises if one studies a \(\gamma_{\mu}\otimes\gamma^{\mu}\)
interaction). We therefore
cannot explicitly exclude a scalar confining potential,
mainly because with the present parameters we were not able to find
stationary solutions for higher angular momenta.
For these reasons we also omitted in the Regge plot the state with
angular momentum j=4.
The \(f_1\) meson mass is larger than the mass of the \(f_2\) meson.
Taken together with the results for the vector case this
indicates that the spin orbit splitting can only
be explained by a mixture of scalar and vector type interaction.

\subsection{Decay Observables}
In this section we will discuss the influence of the parameters
on the decay observables of the pseudoscalar and vector
mesons.

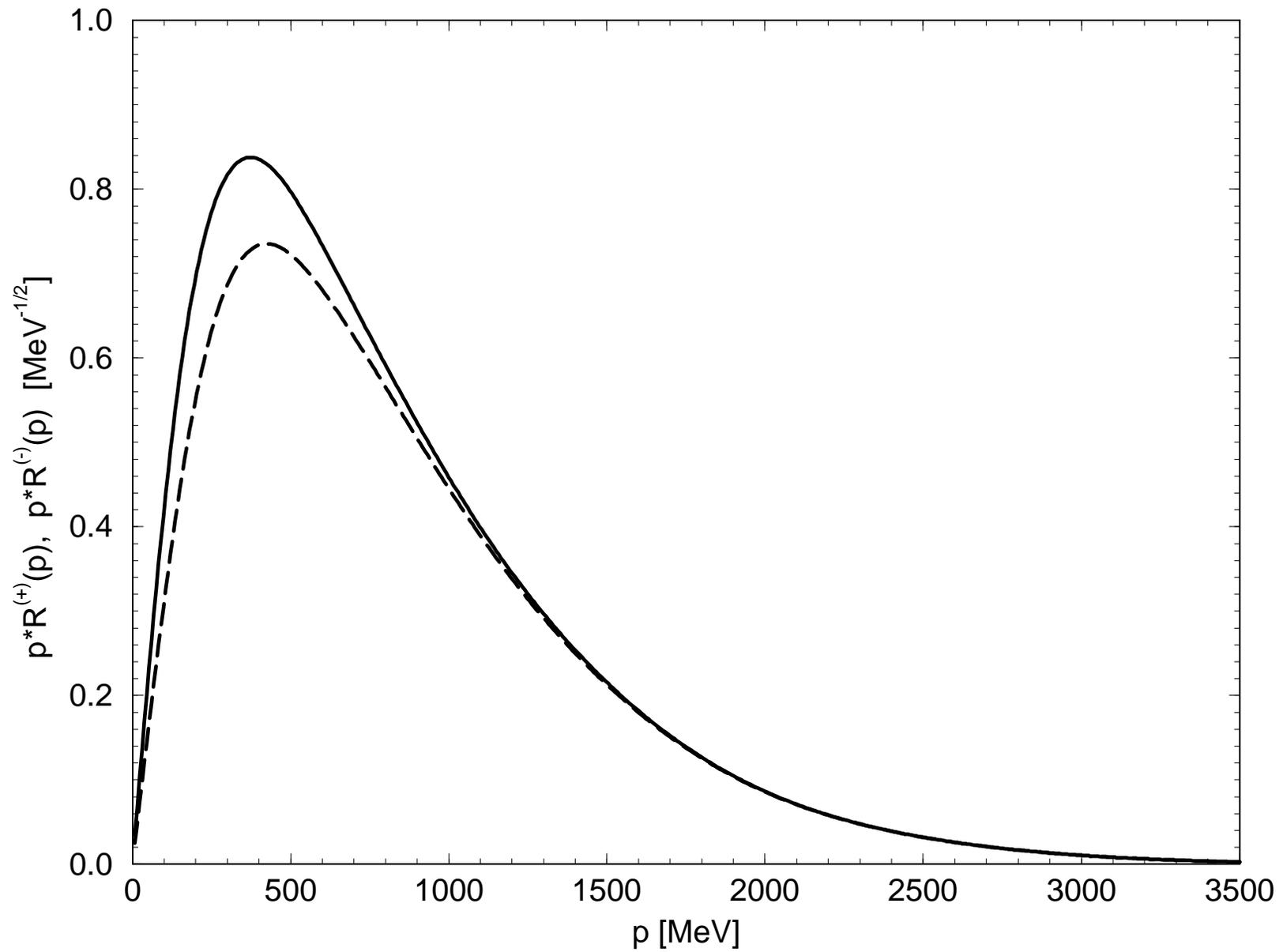
\begin{figure}
\caption{Radial Pion amplitudes
\(p\,{\cal R}^{(+)}_{00}(p)\) (upper component, solid curve) and
\(p\,{\cal R}^{(-)}_{00}(p)\) (lower component, dashed curve)
with the parameters of model V1}
\label{wf1}
\end{figure}

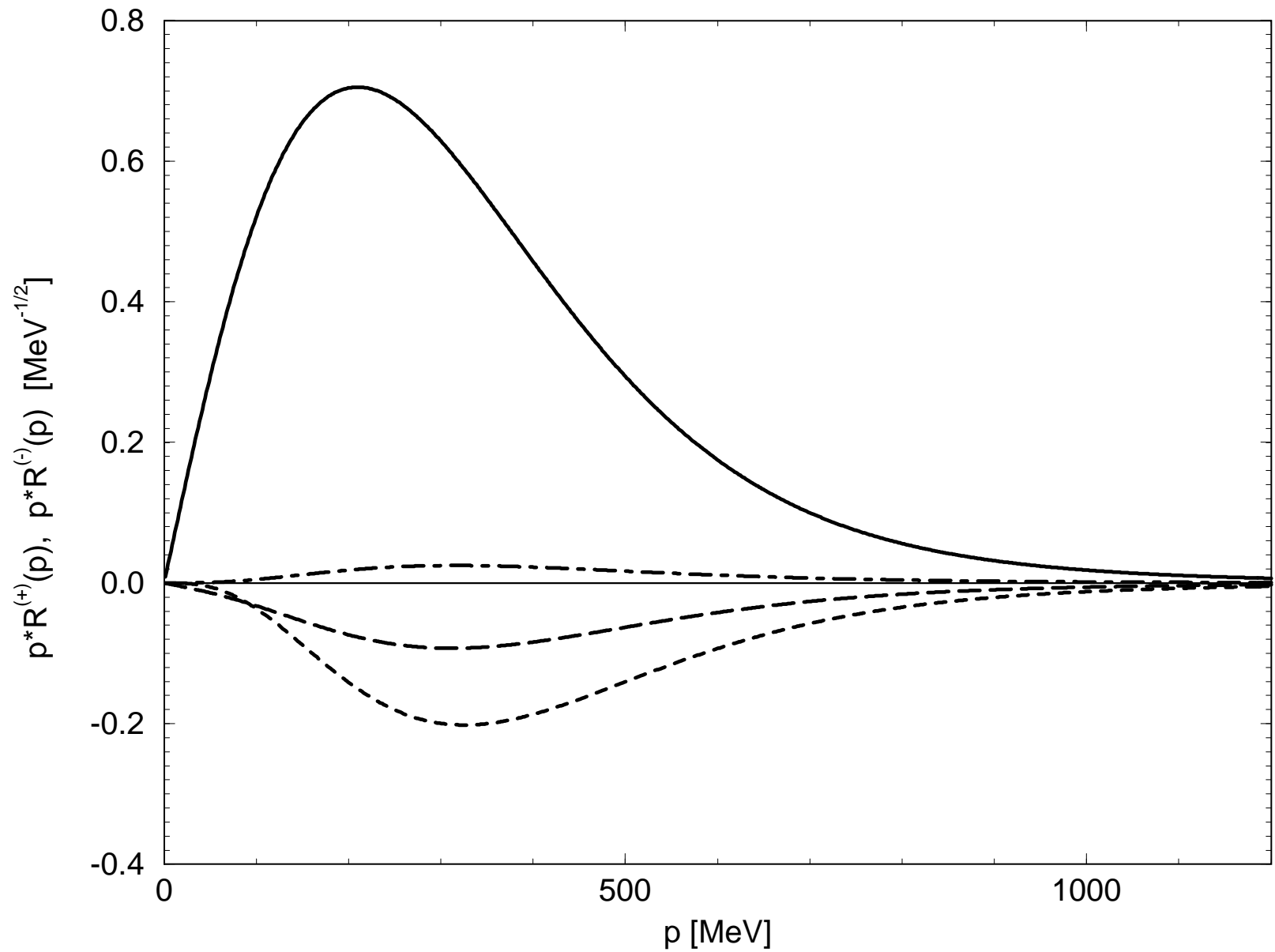
\begin{figure}
\caption{Radial Rho amplitudes
\(p\,{\cal R}^{(+)}_{01}(p)\) (upper s-wave component, solid curve),
\(p\,{\cal R}^{(-)}_{01}(p)\) (lower s-wave component, long dashed curve),
\(p\,{\cal R}^{(+)}_{21}(p)\) (upper d-wave component, dashed dotted curve),
\(p\,{\cal R}^{(-)}_{21}(p)\) (lower d-wave component, short dashed curve)
with the parameters of model V1}
\label{wf2}
\end{figure}

The parameters of Model V1 have been chosen in order
to give a good description of the masses and the
decays of the pseudoscalar and vector ground states.
As can be seen in Tab.\ref{decays}, we obtain an almost
quantitative agreement.
As an important result we consider the fact
that the \( \pi\) and \(\eta \), which usually are interpreted as
Goldstone particles of the of chiral symmetry breaking, also may be
understood as bound states of quark and antiquark, albeit with
relatively small quark masses. This is reflected in the
simultaneous agreement we obtain for the pion decay constant \(f_{\pi}\) and
the decay width \( \pi^0 \rightarrow \gamma\gamma \), which in the
alternative Goldstone picture are related to the Adler Bell Jackiv
anomaly \cite{ABJ}. It becomes clear that a relativistic
treatment of the decay formula and of the normalization are important for a
correct understanding of the pion as deeply bound quark antiquark state.
This can also be seen in Fig.\ref{wf1} showing the upper and lower
component \(\Phi^{++},\,\Phi^{--}\)
of the pion amplitude. In contrast to the wave function
for the \(\rho\) meson (Fig.\ref{wf2}) the upper pion amplitude
is only about 10\% larger than the
lower one in the region of small relative momenta.
For higher momenta the amplitudes even become equal.
This fact obviously leads to important cancellations for the normalization
and for the decay constant.
This was already emphasized in an early
quark model \cite{LS2}, where the effects of
different estimates for the relation between the upper and lower
component on weak decay constants for pion and kaon were analyzed.
For a comparison
with nonrelativistic decay formulas compare Sec.\ref{V1}.
Although the pion amplitude has significant contributions up to
momenta of about
4GeV/c, the main part e.g. of the integral for the \(\pi^0\) decay width
comes from momenta of about 150 MeV/c (as this is the scale of
nonstrange and pion mass).

For the \(\eta \rightarrow \gamma\gamma \) decay we also find
excellent agreement with experiment, whereas the process for the
\(\eta ' \) is underestimated. The results depend strongly on the
correct \(n\bar{n}\)- \(s\bar{s}\) mixing, as e.g. for the \(\eta\)
we obtain a negative interference. The mixing (due to the instanton
induced interaction) can be compared to a simple model given by Rosner
\cite{Ros}. The physical mesons are expanded in a basis of three
states \( |N> =1/\sqrt{2}\,|u\bar{u}+d\bar{d}>\),
\( |S> =|s\bar{s}>\) and \(|G> = |Gluonium>\):
\begin{equation}
   |\eta > = X_{\eta}|N> + Y_{\eta}|S> + Z_{\eta}|G>
\end{equation}
The coefficients may be estimated from electromagnetic transitions
\cite{Bal}. We compared the results for the absolute values
of X and Y in Tab.\ref{mix} with the contributions of the nonstrange
and strange part of the amplitude to the relativistic norm.
The results agree well in the case of the \(\eta\), but not for the
\(\eta '\). Experimental results indicate a larger gluonic component
for the \(\eta '\), which could modify the results.

The leptonic decay widths for the vector mesons are also in good
agreement with the data. This is essentially due to the small
confinement slope, which determines the size of these mesons.
We conclude that a consistent description of all the ground state
pseudoscalar and vector mesons is possible in this framework. However,
more observables like electromagnetic transitions or the pion form factor will
be calculated in the future to substantiate this statement.

The agreement with experimental data for model V2, which was designed
to reproduce the higher resonances and Regge trajectories, is only of
a qualitative character (Tab.\ref{decays}).
Decay constants and photon decay widths disagree by about 50\%,
which is essentially due to the large quark
mass. The leptonic decay widths of the vector mesons are overestimated
due to the steep confinement, which enlarges the amplitudes at the
origin in coordinate space.

Comparing the two parameter sets for vector confinement we find: V1 with light
quark masses gives a quantitative description for the vector
and pseudoscalar ground states, but
only a qualitative picture of the Regge behavior. For V2 with
large quark masses the situation is opposite. This might be an
indication that for large distances the mass of the quarks
effectively should increase due to some additional mass of a
string.

For model S we find good agreement for the vector mesons, but
not as good for the pseudoscalars. Although it is possible to describe
the latter
with a smaller quark mass, we could not obtain a
quantitative adjustment for both \( 0^-\)
and \(1^-\) with a scalar confining kernel.

\subsection{Comparison with nonrelativistic results} \label{V1}
In order to estimate the relevance of relativistic effects in our
model it is useful to compare the results with the corresponding
ones computed in the nonrelativistic quark model.
In the nonrelativistic limit the Salpeter equation reduces
to the usual Schr\"odinger equation with the hamiltonian
given by
\begin{equation}
H = a_c + b_cr + 8\,G_{f_2f_1',f_1f_2'}\,\delta^3(\vec{r})
\end{equation}
where the \(\delta\)-function again has to be regularized according
to eq.(\ref{regtho}). The mass spectra for this hamiltonian have
already been investigated in an earlier work \cite{Bla}
being in good agreement with experiment. In the following
we will use the wave functions \( \psi (\vec{r}) \)
of \cite{Bla} to calculate
the decay observables using the well known formulas
\cite{Bey,vRW,KR}
\begin{eqnarray}
f_{\pi} &=& \frac{2\,\sqrt{3}}{\sqrt{M}}\,|\psi(0)| \\
\Gamma(1^- \rightarrow l^+l^-) &=&
\frac{16\,\pi\,\alpha^2\,\tilde{e}_q^2}{M^2}  \,|\psi(0)|^2 \\
\Gamma(0^- \rightarrow \gamma\gamma) &=&
\frac{12\,\pi\,\alpha^2\,\tilde{e}_q^4}{m_q^2}\,|\psi(0)|^2 \\
\end{eqnarray}
Here \(M\) is the experimental meson mass, \(m_q\) is the quark mass,
\(\alpha=1/137\) and
\(\tilde{e}_q\) gives the quark charge in units of
the proton charge according to the flavor
composition of the meson. The results for these decays
are given in Tab.\ref{decays}.
One finds that leptonic decays can already be described reasonably
in a nonrelativistic framework. On the other hand the weak decay
constants and especially the two photon decay widths are far away
from the experimental data. This discrepancy cannot be cured
by changing the parameters within reasonable limits.
We therefore conclude that a relativistic treatment is essential
for the pseudoscalar mesons.

\subsection{Discussion of the gap equations} \label{V2}
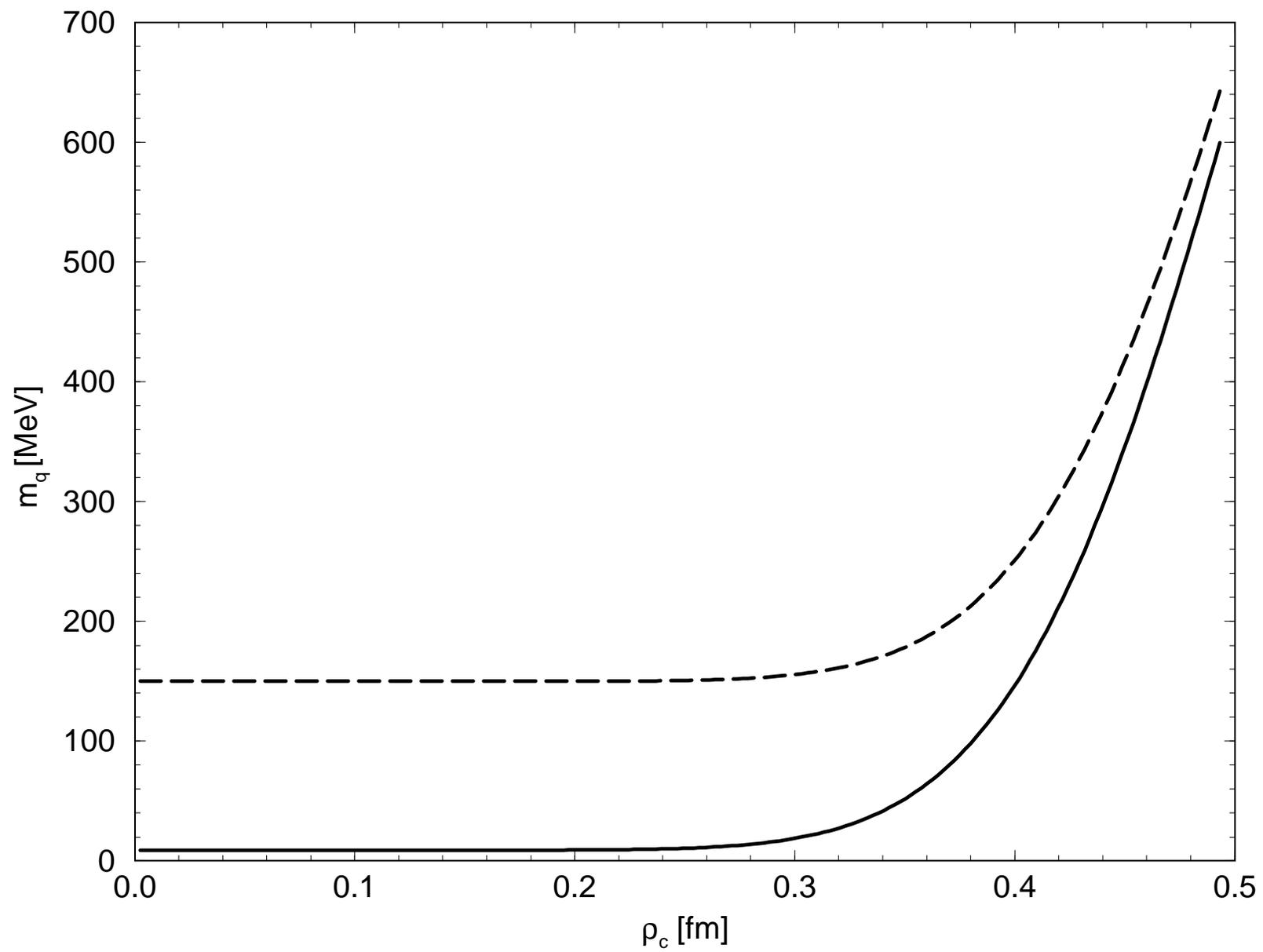
\begin{figure}
\caption{Effective mass dependence for the non strange (solid curve)
  and strange quark (dashed curve) on the instanton size cutoff \(\rho_c\).}
\label{gapplot1}
\end{figure}

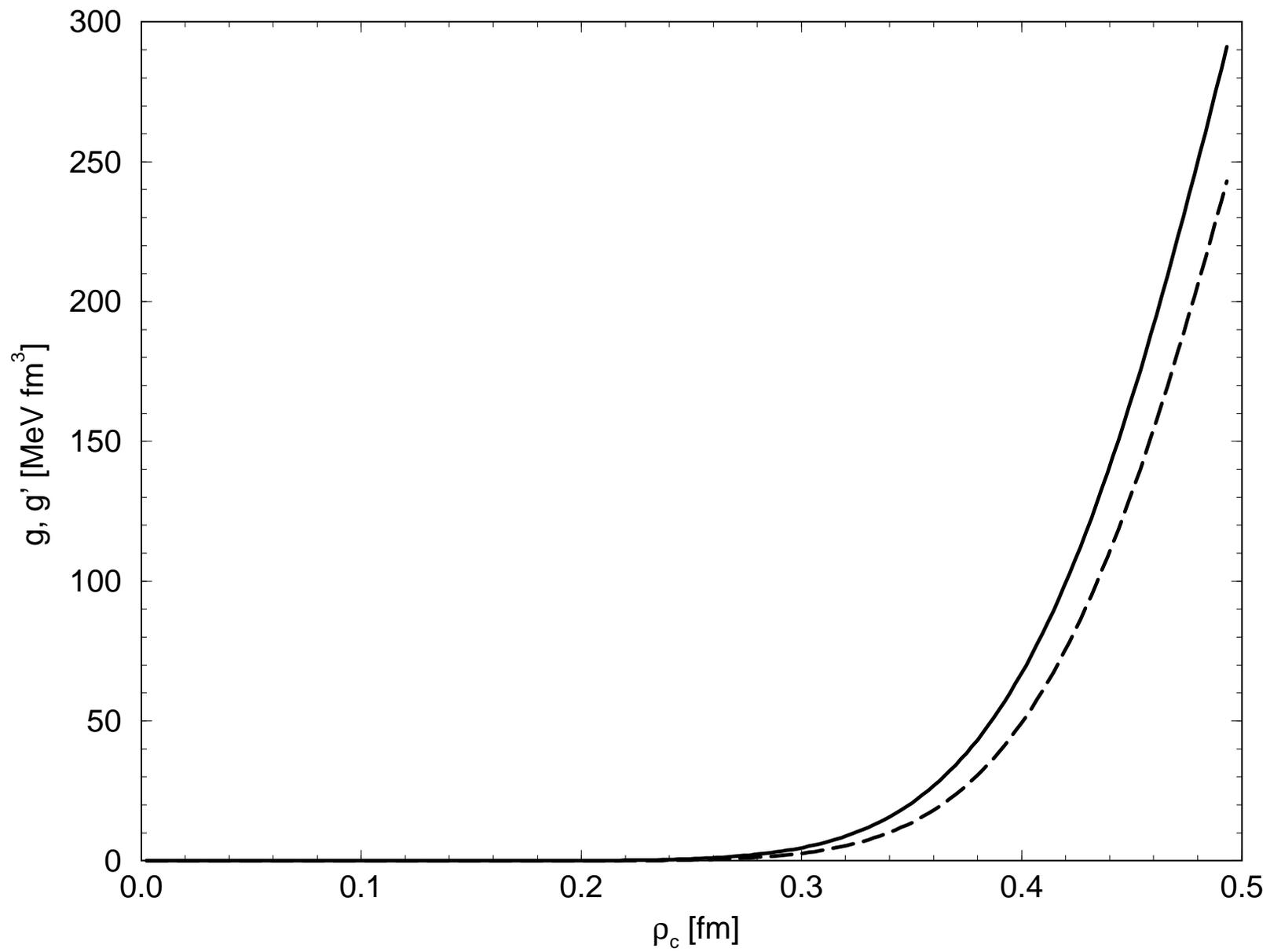
\begin{figure}
\caption{Effective coupling constants for the instanton induced
  non-strange quark interaction \(g\) (solid curve) and the strange
  non-strange quark interaction \(g'\) (dashed curve) as a function of
  the instanton size cutoff \(\rho_c\), see
  eq.(\protect\ref{geff},\protect\ref{ggp})}
\label{gapplot2}
\end{figure}

Due to the process of chiral symmetry breaking the 't Hooft
interaction leads to relations for the effective constituent
quark masses \(m_n,\,m_s\) and the coupling constants
\(g,\,g'\) as given in eqs.(\ref{gapeq}).
In order to check if these relations are qualitatively
compatible with the fitted parameter sets we use
\(\Lambda_{_{QCD}}=200\,MeV\), \(m_n^0=9\,MeV\),
\(m_s^0=150\,MeV\), \(\langle \bar{q}_nq_n \rangle = (-225\,MeV)^3\)
and \(\langle \bar{q}_sq_s \rangle = 0.8\,\langle \bar{q}_nq_n \rangle
\) (compare \cite{RRY}) and plot \(m_n,\,m_s,\,g\) and \(g'\) as
functions of the instanton size cutoff \(\rho_c\) as shown in
Figs.\ref{gapplot1},\ref{gapplot2}.

Because of the delicate dependence on the condensate values
and due to the regularization procedure in the
't Hooft kernel one should not expect quantitative agreement
with the fitted parameter sets.

For \(\rho_c=0.408\,fm\) one finds e.g.
\(m_n=170\,MeV\), \(m_s=270\,MeV\), \(g=79\,MeV\,fm^3\)
and \(g'=58\,MeV\,fm^3\). Despite of the strange quark mass
which comes out too small the other parameters are
quite close to the values of parameter set V1.
Note that \(\rho_c\) is almost equal to the effective
range \(\Lambda\) of the 't Hooft interaction.
Furthermore we find that \(g'\) is smaller than \(g\)
for all values of \(\rho_c\)
as is the case for all three parameter sets.

\section{Summary and conclusion} \label{VI}
Within the framework of the instantaneous Bethe Salpeter equation
we investigated different models for mesons as bound states of quark
and antiquark.
We consider it to be the main result of this work
that the masses, weak decay constants and two photon widths
of the light pseudoscalar mesons (\(\pi,\,\eta,\,K\))
can be understood quantitatively in terms of a \(q\bar{q}\)
description, alternatively to the Goldstone picture. With
the same parameters also the masses and leptonic decays of the vector mesons
can be reproduced.

The pseudoscalar mesons are dominantly affected by an instanton
induced interaction, which apart from the \(\pi,\eta\) splitting
gives the correct \(n\bar{n}-,s\bar{s}-\)mixing for the \(\eta\) meson.
The self energy corrections and coupling constants due to the
resulting chiral symmetry breaking
are compatible with the parameters we used for
the quark antiquark interaction. In contrast to nonrelativistic
calculations instanton effects appear also in the scalar sector and lead to a
isoscalar state with imaginary mass and zero norm.

Concerning the nature of the confinement kernel we find that a
vector type interaction can reproduce the Regge trajectories,
although with a larger quark mass than the one needed to describe the
lowest lying mesons.
This might be an indication that due to string effects
the quark mass should increase with distance.

We were only able to find reliable solutions with a scalar kernel for
relatively large quark masses and weak confinement.
With these parameters, however, one cannot
reproduce quantitatively the ground state mesons or the Regge trajectories.
Nevertheless the spin orbit splitting indicates
the existence of a scalar component in the interaction in order
to cancel the large mass differences coming from the
vector structure.

{\bf Acknowledgments:} We are thankful to M.Fuchs for constant help
in numerical questions.

\begin{appendix}
\section{The 't Hooft interaction} \label{thoapp}
\subsection{The Lagrangian}
As shown by Shifman, Vainshtein and Zakharov \cite{SVZ} the contribution
of an instanton-antiinstanton configuration to the effective quark Lagrangian
for three quark flavors \(u,\,d,\,s\) is given by
\begin{eqnarray}
\lefteqn{
\Delta{\cal L} = \int{} d\rho \frac{d_0(\rho)}{\rho^5}
\Biggl\{
\Biggl[   \prod_{i=1}^3
\left(    m_i^0\rho
        - \frac{4}{3} \pi^2 \rho^3 ( \bar{q}_{iR}q_{iL})
\right) } \nonumber \\ & &
+ \frac{3}{32}
\left( \frac{4}{3} \pi^2 \rho^3
\right)^2
\Biggl[ \Big\{
( \bar{q}_{1R} \lambda^a q_{1L} ) ( \bar{q}_{2R} \lambda^a q_{2L} )
          \nonumber \\ & & -
        \frac{3}{4} ( \bar{q}_{1R} \sigma_{\mu\nu} \lambda^a q_{1L} )
                    ( \bar{q}_{2R} \sigma^{\mu\nu} \lambda^a q_{2L} )
        \Bigr\} \nonumber \\ & &
  \;\;\;  \left( m_3^0\,\rho - \frac{4}{3} \pi^2
  \rho^3(\bar{q}_{3R}q_{3L}) \right)
        \nonumber \\ & &
        + \frac{9}{40} \left( \frac{4}{3} \pi^2 \rho^3 \right) d^{abc}
          ( \bar{q}_{1R} \sigma_{\mu\nu } \lambda^a q_{1L} ) \nonumber \\ &&
  \;\;\;  ( \bar{q}_{2R} \sigma^{\mu\nu } \lambda^b q_{2L} )
          ( \bar{q}_{3R}                  \lambda^c q_{3L} )
        \nonumber\\ & & + \hfill
        \mbox{cycl. perm. of }(123)
        \Biggr]
        \nonumber\\ & &
      + \frac{9}{256} i \left( \frac{4}{3} \pi^2 \rho^3 \right) ^3 f^{abc}
          ( \bar{q}_{1R} \sigma_{\mu}^{\;\nu } \lambda^a q_{1L} )
          \nonumber \\&&
  \;\;\;  ( \bar{q}_{2R} \sigma_{\nu}^{\;\tau} \lambda^b q_{2L} )
          ( \bar{q}_{3R} \sigma_{\tau}^{\;\mu} \lambda^c q_{3L} )
        \nonumber\\ & &
      + \frac{9}{320} \left( \frac{4}{3} \pi^2 \rho^3 \right) ^3 d^{abc}
          ( \bar{q}_{1R} \lambda^a q_{1L} )
          ( \bar{q}_{2R} \lambda^b q_{2L} ) \nonumber \\ &&
  \;\;\;  ( \bar{q}_{3R} \lambda^c q_{3L} )
      \Biggr]
\;\;\;\;
      + (R \leftrightarrow L)
    \Biggr\}
\end{eqnarray}
with $\sigma_{\mu\nu} := \left[ \gamma_{\mu},\gamma_{\nu} \right]/2$
and \( q_{iL} := \frac{1}{2} (1+\gamma_5) q_i\),
 \(\;\;q_{iR} := \frac{1}{2} (1-\gamma_5) q_i\)
being the projections of the quark Dirac operators \(q_i\)
onto left and right handed components.
Furthermore $i=1,2,3=u,d,s$ denotes the flavor degrees of freedom,
$m_i^0$ the corresponding current quark masses,
$\lambda^{a}$ $(a=1,\ldots,8)$ are the color matrices and
$f^{abc}$, $d^{abc}$ are the standard $SU(3)$ structure constants
defined by the commutator
\( [\lambda^a,\lambda^b]_{-} = 2i\, f^{abc}\, \lambda^c \)
and the anticommutator
\( [\lambda^a,\lambda^b]_{+} = \frac{4}{3} \delta^{ab} + 2\, d^{abc}\,
\lambda^c \).
The 't Hooft interaction leads
to spontaneous chiral symmetry breaking as can be seen by normal ordering
\({\cal L}={\cal L}_0+\Delta{\cal L}\)
with respect to the physical QCD vacuum, where
\({\cal L}_0 = \sum_{j=1}^3\,(i\bar{q}_j\gamma^{\mu}\partial_{\mu}q_j
- m_j^0\,\bar{q}q)\) is the free quark Lagrangian.
Using the Wick theorem one finally obtains eq.(\ref{normord}).

\subsection{The interaction between two quarks}
After normal ordering the two body interaction term is given by
\begin{eqnarray}
    \Delta{\cal L}{(2)} & = &  g_{\mbox{eff}}(3)
      \Bigl\{
        \,:\! (\bar{q}_{1R}q_{1L}\bar{q}_{2R}q_{2L})\! :\,
        \nonumber  \\ & & +
        \frac{3}{32}\,:\!
        \Bigl[
           (\bar{q}_{1R} \lambda^{a} q_{1L})
           (\bar{q}_{2R} \lambda^{a} q_{2L})
           \nonumber \\ & & -
           \frac{3}{4} (\bar{q}_{1R} \sigma_{\mu\nu} \lambda^{a} q_{1L})
                       (\bar{q}_{2R} \sigma^{\mu\nu} \lambda^{a} q_{2L})
        \Bigr]\!: \nonumber \\ &&
        + \hfill
        (R \leftrightarrow L) \Bigr\}
      + \mbox{cycl. perm. of } (123)
\end{eqnarray}
with \(g_{\mbox{eff}}(i)\) given in eq.(\ref{geff}).
One can transform the two body force
into a more transparent form
using the notation $\varepsilon_{ijk}, i=u,d,s$ with $\varepsilon_{uds}=1$.
Insert
\(q_{i\,L,R} =  (1 \pm \gamma_5)/2\,q_i \)
and use the relations
\begin{equation}
   \sigma_{\mu \nu} \! \cdot \! \sigma^{\mu \nu}
   + \sigma_{\mu \nu}\gamma_5 \! \cdot \! \sigma^{\mu \nu}\gamma_5  =
   -4\,\left( \Sigma^i \! \cdot \! \Sigma^i
   + \gamma_5\Sigma^i \! \cdot \! \gamma_5\Sigma^i \right)
\end{equation}
with $\Sigma = diag(\sigma,\sigma)$ where the notation
\(
(A \! \cdot \! B)_{ij,kl}=A_{ik}\,B_{jl}
\)
has been used.
Furthermore use
\begin{eqnarray}
  \Sigma^k \cdot \Sigma^k   & = & 2\,\Pi^S - 1^S
  \nonumber \\
  \lambda^a \cdot \lambda^a & = & 2\,\Pi^C - \frac{2}{3}1^C
\end{eqnarray}
with $\Pi^S,\,\Pi^F$ and $\Pi^C$ being exchange operators
in spin, flavor and color defined as
\(
\Pi_{ij,kl} = \delta_{il}\,\delta_{jk}
\)
On the antisymmetric tensors one has
\( \Pi^S\,\Pi^F\,\Pi^C = -1 \) which can be used to eliminate
the spin dependence leading to eq.(\ref{dl2}).

\section{Numerical solution of the Salpeter equation}
\subsection{General framework}
In this section we briefly sketch the
numerical method outlined in ref.\cite{RMMP}.
With the standard Dirac representation (see e.g.\cite{IZ})
\(\Phi\) is a 4\(\times\)4-matrix in spinor space
that can be written in block matrix form as
\begin{equation}
\Phi =
\left( \begin{array}{*{2}{c}}
\Phi^{+-} & \Phi^{++} \\
\Phi^{--} & \Phi^{-+}
\end{array} \right)
 = \hat{\Phi}\,(\Phi^{++},\Phi^{--})
\end{equation}
where each component is a 2\(\times\)2-matrix. The definition
of the bilinear function \(\hat{\Phi}\) is motivated
by the fact that only two of these four
amplitudes are independent since the Salpeter equation implies
\begin{eqnarray}
\Phi^{+-} &=& +c_1 \Phi^{++} s - c_2 s \Phi^{--} \nonumber \\
\Phi^{-+} &=& -c_1 \Phi^{--} s + c_2 s \Phi^{++} \label{pmmp}
\end{eqnarray}
where we use the shorthand notation
\(
s = \vec{\sigma}\vec{p}\,,\;\;
c_i = \omega_i/(\omega_1 m_2 + \omega_2 m_1)
\).

The transformation properties of the Salpeter amplitude imply
that we can decompose
\(\Phi^{++},\,\Phi^{--}\) as
\begin{eqnarray}
\Phi^{++}(\vec{p}) &=&
 \sum_{L\,S}\,{\cal R}_{LS}^{(+)}(p) \,
 \left[Y_L(\Omega_p) \otimes \varphi_S \right]^J \nonumber \\
\Phi^{--}(\vec{p}) &=&
 \sum_{L\,S}\,{\cal R}_{LS}^{(-)}(p) \,
 \left[Y_L(\Omega_p) \otimes \varphi_S \right]^J \label{ang}
\end{eqnarray}
with the spin matrix
\( (\varphi_{Sq}\,i\sigma_2)_{mm'}=\langle 1/2\,m\,1/2\,m'|S\,q \rangle\).
\(L,\,S\) have to respect the usual constraints coming from
parity and charge conjugation.
Let
\begin{equation}
 E_i(\vec{p}) = R_{n_i L_i}(p)\,
 \left[ Y_{L_i}(\Omega_p) \otimes \varphi_{S_i} \right]^J_{M_J} \label{80}
\end{equation}
be a complete set of 2\(\times\)2 basis functions orthonormal with respect to
the scalar product given by
\(
 \left( E_i \right.\left|\,E_j \right) =
\int \,\mbox{tr}\,
\left[E_i^{\dagger}(\vec{p})\,E_j(\vec{p})\right]
= \delta_{ij}
\).
These basis functions can be used to expand \(\Phi^{++}\) and \(\Phi^{--}\)
with corresponding (usually real) expansion coefficients \(a_i^{(+)}\)
and \(a_i^{(-)}\).
Inserting these expansions into \(\hat{\Phi}\) gives
\begin{equation}
\psi = \sum_{i=1}^{\infty}\,
\left(a_i^{(+)}\,e_i^{(+)} + a_i^{(-)}\,e_i^{(-)} \right)
\end{equation}
with
\(e_i^{(+)} = \hat{\Phi}\,(E_i,0)\,\gamma^0\) and
\(e_i^{(-)} = \hat{\Phi}\,(0,E_i)\,\gamma^0\).
Defining the matrix elements (which are real in our case)
\(H^{ss'}_{ij} = \langle e_i^{(s)} | {\cal H}\,e_j^{(s')} \rangle \)
and
\(N^{ss'}_{ij} = \langle e_i^{(s)} | e_j^{(s')} \rangle \)
the Salpeter equation \( {\cal H}\psi = M\psi \)
can now be written in the form of a matrix equation as
\begin{equation}
\left( \!\! \begin{array}{*{2}{c}}
H^{++} & H^{+-} \\
H^{+-} & H^{++}
\end{array} \!\!\right)
\left( \!\!\begin{array}{*{1}{c}}
a^{(+)} \\ a^{(-)}
\end{array} \!\!\right)
= M\,
\left( \!\!\begin{array}{*{2}{c}}
N^{++} & 0 \\
0      & -N^{++}
\end{array} \!\!\right)
\left( \!\!\begin{array}{*{1}{c}}
a^{(+)} \\ a^{(-)}
\end{array} \!\!\right) \label{matrix}
\end{equation}
We solve this equation within a finite basis \(i \le i_{max}=10\)
and use the variational principle \(\delta M = 0\) looking
for stationary points of \(M\) as a function of \(\beta\),
where \(\beta\) is a variational parameter with the physical
dimension \(MeV^{-1}\) that sets
the absolute scale of the basis functions as
\(E_i^{\beta}(\vec{p})=\beta^{3/2}\,E_i^{\beta=1}(\vec{p}\beta)\).
For the basis functions in momentum space the following
functions have been used:
\begin{equation}
R_{nL}(y)=N_{nL}\,y^L\,L_n^{2L+2}(y)\,e^{-y/2}
\end{equation} with
\(N_{nL}\) being a normalization coefficient,
\(y=p\beta\) and \(L_n^{2L+2}(y)\) being a Laguerre polynomial.
The fouriertransformed basis states also can be given
in analytical form.
About ten basis states are sufficient to solve
the Salpeter equation with rather high accuracy.
The usual choice of \(3\)-dimensional harmonic oscillator functions
is less favored here since their asymptotic behavior
\(\sim e^{-y^2/2}\) for \(y\rightarrow\infty\) turns out
to be not appropriate for deeply bound
states like the pion. For other mesons, however, we have checked that
both basis systems give equal results within numerical errors.

\subsection{Expectation values}
In order to compute the matrix elements present in eq.(\ref{matrix})
it is useful to rewrite \(\langle \psi|\psi \rangle\) and
\(\langle \psi|{\cal H}\psi \rangle\) in terms of \(\Phi^{++}\)
and \(\Phi^{--}\) using eq.(\ref{pmmp}). We find
\begin{eqnarray}
\langle \psi|\psi \rangle
&=& \int\,\frac{2\,\omega_1\,\omega_2}
{\omega_1\,m_2+\omega_2\,m_1} \nonumber \\ &&
\mbox{tr}\,\left( (\Phi^{++})^{\dagger}\Phi^{++}
- (\Phi^{--})^{\dagger}\Phi^{--}\right) \label{normexp}
\end{eqnarray}
for the norm. The matrix elements of \({\cal H}\) can be split as
\begin{eqnarray}
\langle \psi| {\cal H}\psi \rangle &=&
\langle \psi| {\cal T}\psi \rangle +
\langle \psi| {\cal V}\psi \rangle \label{R1a} \\
\langle \psi| {\cal T}\psi \rangle
  & = & \int\,(\omega_1+\omega_2) \,
\mbox{tr}\,\left[\Phi^{\dagger}\,\Phi\right] \label{R1b} \\
\langle \psi| {\cal V}\psi \rangle
  & = & -\int \!\int'\,
\mbox{tr}\,\left[
\Phi^{\dagger}\,\gamma^0\,(V\Phi')
\,\gamma^0 \right] \label{R1c}
\end{eqnarray}
with \(\int = \int d^3p/(2\pi)^3\).
The basic formula for the calculation of the kinetic energy
and confinement matrix elements is given by
\begin{eqnarray}
\lefteqn{\mbox{tr}\,(\Phi^{\dagger}\Phi') = } \label{R2} \\
& = \;\mbox{tr} &
[
(\Phi^{++})^{\dagger} (\Phi^{++})' +
(\Phi^{--})^{\dagger} (\Phi^{--})' + \nonumber \\
&& +
(\Phi^{+-})^{\dagger} (\Phi^{+-})' +
(\Phi^{-+})^{\dagger} (\Phi^{-+})' ] = \nonumber \\
& = \;\mbox{tr} &
[
(\Phi^{++})^{\dagger} (\Phi^{++})' +
(\Phi^{--})^{\dagger} (\Phi^{--})' + \nonumber\\
&& +
c_1c_1'\,(\Phi^{++})^{\dagger} (\Phi^{++})'\,s's -
c_1c_2'\,(\Phi^{++})^{\dagger} \,s'(\Phi^{--})'s - \nonumber \\
&& -
c_2c_1'\,(\Phi^{--})^{\dagger} \,s(\Phi^{++})'\,s' +
c_2c_2'\,(\Phi^{--})^{\dagger} \,ss'(\Phi^{--})' + \nonumber \\
&& +
c_1c_1'\,(\Phi^{--})^{\dagger} (\Phi^{--})'\,s's -
c_1c_2'\,(\Phi^{--})^{\dagger} \,s'(\Phi^{++})'s - \nonumber \\
&& -
c_2c_1'\,(\Phi^{++})^{\dagger} \,s(\Phi^{--})'\,s' +
c_2c_2'\,(\Phi^{++})^{\dagger} \,ss'(\Phi^{++})' ] \nonumber
\end{eqnarray}
where the prime indicates the dependence on \(\vec{p}\,'\).
To compute the kinetic energy term eq.(\ref{R1b}) set
\(\vec{p}\,'=\vec{p}\) in this equation.

For the interaction term eq.(\ref{R1c}) we investigate the
following contributions:

\subsubsection{Scalar confinement}
{}From the scalar confining kernel eq.(\ref{scalar}) one has
\begin{equation}
\langle \psi|{\cal V}_C^S \,\psi\rangle =
 -\int \! \int'\,{\cal V}_C\,
\mbox{tr}\,\left[
\Phi^{\dagger}\,\gamma^0\,\Phi'
\,\gamma^0 \right] \end{equation}
with \({\cal V}_C = {\cal V}_C((\vec{p}-\vec{p}\,')^2)\).
Since
\begin{equation} \gamma^0\,\Phi\,\gamma^0 =
\left( \begin{array}{*{2}{c}}
 \Phi^{+-} & -\Phi^{++} \\
-\Phi^{--} &  \Phi^{-+}
\end{array} \right)
\end{equation}
the expression for
 \( \mbox{tr}\,\left[
\Phi^{\dagger}\,\gamma^0\,\Phi'
\,\gamma^0 \right] \) is obtained by changing the sign of the
first two terms
 \((\Phi^{++})^{\dagger} (\Phi^{++})'\) and
 \((\Phi^{--})^{\dagger} (\Phi^{--})'\)
in eq.(\ref{R2}).

\subsubsection{Vector confinement}
{}From the vector confining kernel eq.(\ref{vector}) one has
\begin{equation}
\langle \psi| {\cal V}_C^V \,\psi \rangle =
 \int\int'\,{\cal V}_C \,
\mbox{tr}\,\left[
\Phi^{\dagger}\Phi'\right] \end{equation}
and eq.(\ref{R2}) can be applied directly.
Note that a covariant form to write the vector
confinement kernel is
\begin{eqnarray}
\lefteqn{ \left[ K(P,p,p')\,\chi(p')\right] = } \\ &=&
-{\cal V}_C((p_{\perp}-p_{\perp}')^2)\,\frac{1}{P^2}\,
P^{\mu}\gamma_{\mu}\,\chi(p')\,P^{\nu}\gamma_{\nu} \nonumber
\end{eqnarray}
with \(p_{\perp}=p-(Pp)/P^2\),
so that the relation
\begin{equation} P^{\mu}\frac{d}{dP^{\mu}}\,[K(P,p,p')] = 0 \end{equation}
holds as is required to rewrite the normalization condition
for \(\Phi\).

\subsubsection{'t Hooft interaction}
{}From the 't Hooft kernel eq.(\ref{thokern}) we find
(omitting flavor indices)
\begin{eqnarray}
\lefteqn{ \langle \psi |{\cal V}_T\,\psi \rangle = } \\
&=& -4G\,\int\,\int'\,
\mbox{tr}\,\left[
\Phi^{\dagger}\,\gamma^0\,
\left( 1_4\,\mbox{tr}\,\Phi' +
\gamma^5\,\mbox{tr}\,(\Phi'\,\gamma^5)
\right)\,\gamma^0 \right] = \nonumber \\
&=& -4G\,\int\int'\,\left[
\left( \mbox{tr}\,\Phi  \right)^*\,
\left( \mbox{tr}\,\Phi' \right) -
\left( \mbox{tr}\,\Phi  \,\gamma^5 \right)^*\,
\left( \mbox{tr}\,\Phi' \,\gamma^5 \right)
\right] \nonumber
\end{eqnarray}
with
\begin{eqnarray}
\mbox{tr}\,\Phi            &=& \mbox{tr}\,(\Phi^{+-}+\Phi^{-+}) \nonumber \\
\mbox{tr}\,(\Phi \gamma^5) &=& \mbox{tr}\,(\Phi^{++}+\Phi^{--})
\end{eqnarray}
Using the decomposition eq.(\ref{ang}) and
\( \mbox{tr}\,\varphi_{S\,q} = \sqrt{2}\,\delta_{S\,0}\)
one obtains
\begin{eqnarray}
\mbox{tr}\,\Phi^{++}_{JM_J}(\vec{p}) &=&
\sqrt{2}\,{\cal R}_{NJ0}^{(+)}(p)\,Y_{JM_J}(\Omega_p)
\,\delta_{S\,0} \\
\mbox{tr}\,\Phi^{--}_{JM_J}(\vec{p}) &=&
\sqrt{2}\,{\cal R}_{NJ0}^{(-)}(p)\,Y_{JM_J}(\Omega_p)
\,\delta_{S\,0}
\end{eqnarray}
We further see that
\begin{equation} \mbox{tr}\,(\varphi_{S\,q}\,\vec{\sigma}\vec{p})
 = \mbox{tr}\,(\vec{\sigma}\vec{p}\,\varphi_{S\,q})
 = \sqrt{2}\,\delta_{S\,1}\,p_{q} \end{equation}
with the spherical components
\(p_q = \sqrt{4\pi/3}\; p\,Y_{1\,q}(\Omega_p)\).
We find
\begin{eqnarray}
\lefteqn{\int d\Omega_p\,\mbox{tr}\,(\Phi \gamma^5) = } \\
&=& \sqrt{2}\,\sqrt{4\pi}\,
  ({\cal R}_{NJ0}^{(+)}(p)
 + {\cal R}_{NJ0}^{(-)}(p))\,
\delta_{S\,0}\,\delta_{L\,0}\,\delta_{J\,0} \nonumber \\
\lefteqn{\int d\Omega_p\,\mbox{tr}\,\Phi = } \\
&=& -\sqrt{2}\,\sqrt{4\pi}\,
p\,(c_1+c_2)\,
  ({\cal R}_{NJ0}^{(+)}(p)
 - {\cal R}_{NJ0}^{(-)}(p))\,
\delta_{S\,1}\,\delta_{L\,1}\,\delta_{J\,0} \nonumber
\end{eqnarray}

\subsection{Matrix elements}
The expressions obtained above can now be used to compute
the required matrix elements. In order to calculate e.g.
\(N^{++}_{ij}=\langle e_i^{(+)}|e_j^{(+)} \rangle\)
one has to replace in eq.(\ref{normexp}) \((\Phi^{++})^{\dagger}\)
by \(E_i^{\dagger}\) and \(\Phi^{++}\) by \(E_j\) setting
\((\Phi^{--})^{\dagger}=0\) and \(\Phi^{--}=0\).
In the following we will use the notation
\begin{equation} (i|f(p)|j) = \int \frac{p^2\,dp}{(2\pi)^3}
\,R_{n_iL_i}(p)\,R_{n_jL_j}(p)\,f(p) \end{equation}
for the radial integrals between the basis states.
These integrals can be effectively computed numerically using e.g.
Gauss quadrature routines.
According to eq.(\ref{R1a}) we write
\(H_{ij}^{ss'}=T_{ij}^{ss'}+V_{ij}^{ss'}\) with
\(V_{ij}^{ss'}=(V_C)_{ij}^{ss'}+(V_T)_{ij}^{ss'}\).

\subsubsection{Normalization matrix elements}
For the normalization matrix elements we find
from eq.(\ref{normexp})
\begin{eqnarray}
N^{++}_{ij} &=& -N^{--}_{ij} =
\left( i \left| \frac{2\omega_1\omega_2}{\omega_1m_2+\omega_2m_1}
\right| j \right) \\
N^{+-}_{ij} &=& N^{-+}_{ij} = 0
\end{eqnarray}

\subsubsection{Kinetic energy matrix elements}
In order to evaluate the angular momentum structure
of matrix elements the relation
\( (-i\sigma_2)\,\vec{\sigma}\,(i\sigma_2)
= -{}^t \vec{\sigma} \label{50} \) is quite useful.
Define the \(q\bar{q}\) spin matrix as
\(\chi_{S\,q}=\varphi_{S\,q}\,i\sigma_2\) and use \(s=\vec{\sigma}\vec{p}\).
Then one can write e.g.
\begin{eqnarray}
\lefteqn{\mbox{tr}\,(\varphi^{\dagger}_{S_i\,M_i}\,s\,\varphi_{S_j\,M_j}\,s)}
\nonumber \\
&=& -\mbox{tr}\,(\chi_{S_i\,M_i}^{\dagger}
\,(s_1 \otimes s_2)\,\chi_{S_j\,M_j}) =
\nonumber \\
&=:& -\left\langle S_i M_i \left| s_1 \otimes s_2 \right| S_j M_j \right\rangle
\end{eqnarray}
with \(s_i=2\vec{S}_i\vec{p}\) where \(\vec{S}_1\)
is the spin operator acting on the first quark and
\(\vec{S}_2\) on the second quark.
It is useful to define
\begin{eqnarray}
S_1(L',S',L,S,J) &:=&
\left\langle [L' \otimes S']^J \left|
 s_1/p \right| [L \otimes S]^J \right\rangle\\
S_2(L',S',L,S,J) &:=&
\left\langle [L' \otimes S']^J \left|
 s_2/p \right| [L \otimes S]^J \right\rangle \nonumber \\
S_{12}(L',S',L,S,J) &:=&
\left\langle [L' \otimes S']^J \left|
 (s_1 \otimes s_2)/p^2 \right| [L \otimes S]^J \right\rangle
\nonumber
\end{eqnarray}
Using the Wigner-Eckart theorem and the notation
\(\hat{L}=\sqrt{2L+1}\) one has
\begin{eqnarray}
\lefteqn{S_1(L',S',L,S,J) = (-1)^{S+S'}\,S_2(L',S',L,S,J) =} \nonumber \\
&=& (-1)^{L'+1}\,\hat{J}\hat{L'}\hat{L}\hat{S'}\hat{S}\,
\sqrt{9}\sqrt{12} \,
\left( \begin{array}{*{3}{c}}
L'& 1 & L \\
0 & 0 & 0
\end{array} \right)
\nonumber \\ && \cdot
\left\{ \begin{array}{*{3}{c}}
L' & S' & J \\
L  & S  & J \\
1  & 1  & 0
\end{array} \right\} \,
\left( \begin{array}{*{3}{c}}
1/2 & 1/2 & S' \\
1/2 & 1/2 & S  \\
1   & 0   & 1
\end{array} \right) \\
\lefteqn{S_{12}(L',S',L,S,J) =} \nonumber \\
&=& \sum_{L_k\,S_k}\;S_1(L',S',L_k,S_k,J)\;S_2(L_k,S_k,L,S,J)
\end{eqnarray}
The kinetic energy matrix elements are now given by
\begin{eqnarray}
\lefteqn{T_{ij}^{++} = T_{ij}^{--} =} \nonumber \\ &=&
(i|(\omega_1+\omega_2)\,(1+p^2(c_1^2+c_2^2))\,|j)
\,\delta_{L_iL_j}\,\delta_{S_iS_j} \nonumber \\
\lefteqn{T_{ij}^{+-} = T_{ij}^{-+} =} \nonumber \\ &=&
(i|(\omega_1+\omega_2)\,2\,p^2\,c_1\,c_2|j)
\,S_{12}(i,j)
\end{eqnarray}
with \(S_{12}(i,j)=S_{12}(L_i,S_i,L_j,S_j,J)\).

\subsubsection{Scalar and vector confinement matrix elements}
The confinement matrix elements can be computed
by inserting two complete sets of basis functions like
\begin{eqnarray}
\lefteqn{\langle i | f_1(\vec{p})\,V(r)\,f_2(\vec{p}\,') | j \rangle = }\\
&=& \sum_{g,h}\,
\langle i | f_1(\vec{p}) | g \rangle \,
\langle g | V(r) | h \rangle \,
\langle h | f_2(\vec{p}\,') | j \rangle \nonumber
\end{eqnarray}
where the matrix element of \(V\) can be evaluated in
coordinate space.
One finds for the scalar confinement matrix elements
\begin{eqnarray}
\lefteqn{ (V_C^S)^{++}_{ij} = (V_C^S)^{--}_{ij} =
(i|{\cal V}_C|j) -\sum_{g,h}} \\
&& \Big\{\;
(i|pc_1|g)\,S_2(i,g)\;
(g|{\cal V}_C |h)\;
(h|pc_1|j)\,S_2(h,j) \nonumber\\
&& +
(i|pc_2|g)\,S_1(i,g)\;
(g|{\cal V}_C |h)\;
(h|pc_2|j)\,S_1(h,j) \;\Big\} \nonumber
\end{eqnarray}
and
\begin{eqnarray}
\lefteqn{ (V_C^S)^{+-}_{ij} = (V_C^S)^{-+}_{ij} = -\sum_{g,h}} \\
&& \Big\{\;
(i|pc_1|g)\,S_2(i,g)\;
(g|{\cal V}_C |h)\;
(h|pc_2|j)\,S_1(h,j) \nonumber\\
&& +
(i|pc_2|g)\,S_1(i,g)\;
(g|{\cal V}_C |h)\;
(h|pc_1|j)\,S_2(h,j) \;\Big\} \nonumber
\end{eqnarray}
with \(\; (g|{\cal V}_C|h)=\int r^2\,dr\,
R^F_{n_iL_i}(r)\,R^F_{n_jL_j}(r)\;{\cal V}_C^F(r)\,
\delta_{L_iL_j}\,\delta_{S_iS_j} \;\) where \(R^F\)
denotes the Fourier transformed basis functions
and \({\cal V}_C^F(r)=a_c+b_cr\).

For the vector confinement matrix elements one only
has to change the sign of \(\sum_{g,h}\) in the two
equations above.

\subsubsection{'t Hooft matrix elements}
The 't Hooft matrix elements with the regularizing potential
can be computed analogously to the confinement matrix elements.
In the unregularized case \(\Lambda \rightarrow 0\)
corresponding to \({\cal V}_{\mbox{reg}}^F(r) \rightarrow \delta(\vec{r})\)
only \(L=0\) basis states contribute to \( (g|{\cal V}_{\mbox{reg}}|h)\).
Consistently only \(L=0\) states will be taken into account
for \( (g|{\cal V}_{\mbox{reg}}|h)\) also for \(\Lambda > 0\)
so that the angular selection rules
are not changed by the regularization.
The result for \(L=S=0\) reads
\begin{eqnarray}
\lefteqn{  (V_T^{++})^{S=0}_{ij} = (V_T^{--})^{S=0}_{ij}
         = (V_T^{+-})^{S=0}_{ij} = (V_T^{-+})^{S=0}_{ij} } \nonumber \\
 & = & 8G\,(i|{\cal V}_{\mbox{reg}}|j) \;
       \delta_{S_i\,0}\,\delta_{S_j\,0}\,\delta_{L_i\,0}\,
       \delta_{L_j\,0}\,\delta_{J\,0}
\end{eqnarray}
and for \(L=S=1\)
\begin{eqnarray}
\lefteqn{  (V_T^{++})^{S=1}_{ij} = (V_T^{--})^{S=1}_{ij}
         = -(V_T^{+-})^{S=1}_{ij} = -(V_T^{-+})^{S=1}_{ij} } \nonumber \\
 & = & -8G\,\sum_{g,h}\,
(i|p\,(c_1+c_2)|g)\,
(g|{\cal V}_{\mbox{reg}}|h)
(h|p\,(c_1+c_2)|j) \nonumber \\ & &
\delta_{S_i\,1}\,\delta_{S_j\,1}\,\delta_{L_i\,1}\,
       \delta_{L_j\,1}\,\delta_{J\,0}
\end{eqnarray}
This result shows
that the 't Hooft interactions affects only mesons
with \(J=0\) and \(L=S=0\) (i.e. pseudoscalar mesons
with \(J^{\pi_P}=0^-\)) or
\(L=S=1\) (i.e. scalar mesons
with \(J^{\pi_P}=0^+\)). In the nonrelativistic limit
the contributions to the scalar mesons vanish.

\end{appendix}

\noindent

\noindent

\begin{table}
\caption{Flavor dependence of the instanton induced interaction
  \(G_{f_2f_3,f_1f_4}\) (see eq.(\protect\ref{gff}))}
\label{tab1}
\begin{tabular}{cccccccccc}
\(f_1f_2 \,\rightarrow\)
        & $u\bar{d}$ &$d\bar{u}$
        & $u\bar{s}$ &$d\bar{s}$ & $s\bar{d}$ & $s\bar{u}$
        & $u\bar{u}$ &$d\bar{d}$ & $s\bar{s}$ \\
$f_3f_4 \,\downarrow$
               &  &  &   &   &   &   &  &  &  \\
\hline
     $u\bar{d}$ &-g& 0&   &   &   &   &  &  &  \\
     $d\bar{u}$ & 0&-g&   &   &   &   &  &  &  \\
\hline
     $u\bar{s}$ &  &  &-g'&  0&  0&  0&  &  &  \\
     $d\bar{s}$ &  &  &  0&-g'&  0&  0&  &  &  \\
     $s\bar{d}$ &  &  &  0&  0&-g'&  0&  &  &  \\
     $s\bar{u}$ &  &  &  0&  0&  0&-g'&  &  &  \\
\hline
     $u\bar{u}$ &  &  &   &   &   &   & 0& g&g' \\
     $d\bar{d}$ &  &  &   &   &   &   & g& 0&g' \\
     $s\bar{s}$ &  &  &   &   &   &   &g'&g'& 0 \\
\end{tabular}
\end{table}

\begin{table}
\caption{Flavor matrix elements of \(G_{f_2f_3,f_1f_4}\) for
  pseudoscalar mesons}
\label{tab2}
\begin{tabular}{cccc}
 & $\pi^0$ & $\eta_n$ & $\eta_s$ \\
\hline
 $\pi^0$  & -g&  0&  0\\
 $\eta_n$ &  0&  g&$\sqrt{2}\,$g' \\
 $\eta_s$ &  0& $\sqrt{2}\,$g'& 0 \\
\end{tabular}
\end{table}

\begin{table}
  \caption{Parameters of the different models (see Sec.\protect\ref{MoPa})}
  \label{param}
  \centering
   \begin{tabular}{ccccc}
       Parameter                &   V1  & V2    & S      \\
     \hline
     \(m_n\)[MeV]               &  170  &  340  & 340    \\
     \(m_s\) [MeV]              &  390  &  568  & 487    \\
     \(a_c\) [MeV]              & -552  & -1340 & -998   \\
     \(b_c\) [MeV/fm]           &  570  &  1400 & 1000   \\
      g  [MeV\,fm\(^3\)]        & 51.67 & 34.65 & 44.79  \\
      g' [MeV\,fm\(^3\)]        & 46.92 & 30.84 & 41.01  \\
     \(\Lambda\) [fm]           & 0.42  &  0.42 & 0.42   \\
\end{tabular}
\end{table}

\begin{table}
  \caption{Spin orbit splitting for the first positive parity mesons
    and the effect of the instanton induced interaction on the
    \(0^{++}\) mesons, where a \(^*\) denotes the
    existence of an additional state with imaginary mass and zero
    norm, see Sec.\protect\ref{Msspc} (all Masses in MeV).}
  \label{spor}
  \centering
   \begin{tabular}{ccccc}
   Meson &  \(J^{PC}\) I  &  V1 &  V2 &  S \\
\hline
   \(a_0 (980)\)   &  \( 0^{++}\)  1  &  960  & 1130 & 1260 \\
   \(f_0 (975)\)   &  \( 0^{++}\)  0  &  950\(^*\) & 1270\(^*\)&  950 \\
   \(f_1 (1285)\)  &  \( 1^{++}\)  0  &  930  & 1060 & 1150 \\
   \(f_2 (1270)\)  &  \( 2^{++}\)  0  &  1100 & 1280 & 1010 \\
\end{tabular}
\end{table}

\begin{table}
  \caption{Comparison of experimental and calculated meson decay
observables for the Salpeter
models V1, V2, S and nonrelativistic results NR}
  \label{decays}
  \centering
   \begin{tabular}{cccccc}
       Mesonic decay  &  experimental \cite{PDG} &  V1  &  V2   &  S  & NR
\\
     \hline
          \(f_{\pi}\) [MeV]
                  & 131.7\(\pm\) 0.2   & 130  &  260  & 200 & 1440    \\
          \( f_K\) [MeV]
                  & 160.6 \(\pm\) 1.4  &  180 &  300  & 210 & 730    \\
          \(\Gamma(\pi^0 \rightarrow \gamma\gamma)\) [eV]
                  & 7.8 \(\pm\) 0.5      & 7.6  &  4.0  & 4.4 & 30000    \\
          \(\Gamma(\eta \rightarrow \gamma\gamma)\) [eV]
                  & 460  \(\pm\) 5       & 440  & 220   & 220 & 18500    \\
          \(\Gamma(\eta' \rightarrow \gamma\gamma)\) [eV]
                  & 4510 \(\pm\) 260     & 2900 &  2030 & 1390& 750    \\
          \(\Gamma(\rho \rightarrow e^+e^-)\) [keV]
                  & 6.8 \(\pm\) 0.3     & 6.8  &  28   & 8.1 & 8.95    \\
          \(\Gamma(\omega \rightarrow e^+e^-)\) [keV]
                  & 0.60 \(\pm\) 0.02   & 0.73 &  3.1  & 0.87& 0.96    \\
          \(\Gamma(\phi \rightarrow e^+e^-)\) [keV]
                  & 1.37 \(\pm\) 0.05  & 1.24 &  4.5  & 1.50& 2.06    \\
\end{tabular}
\end{table}

\begin{table}
  \caption{\(\eta,\eta'\) mixing parameters from BS norm (see
    eq.(\protect\ref{scprd})) compared to
           data calculated from experimental J/\(\Psi\) decays
\protect\cite{Bal}}
  \label{mix}
  \centering
   \begin{tabular}{cccccc}
   Meson &  mixing coefficient   &  J/\(\Psi\) decay & V1 & V2 &  S \\
\hline
   \(\eta (547)\) & \(|X_{\eta}|\)
              & 0.63\(\pm\)0.06 & 0.71 & 0.71 & 0.70\\
                 & \(|Y_{\eta}|\)
              & 0.83\(\pm\)0.13 & 0.70 & 0.70 & 0.72\\
   \(\eta' (958)\) & \(|X_{\eta'}|\)
              & 0.36\(\pm\)0.05 & 0.85 & 0.78 & 0.83\\
                 & \(| Y_{\eta'}|\)
              & 0.72\(\pm\)0.12 & 0.52 & 0.63 & 0.55\\
\end{tabular}
\end{table}

\end{document}